\documentclass[11pt]{article}
\usepackage{makeidx,graphicx}
\begin{document}

\title{Quantum Trajectories, Real, Surreal or an Approximation to a Deeper Process?}
\author{B. J. Hiley, R. E. Callaghan and O. J. E. Maroney.}
\date{Theoretical Physics Research Unit, Birkbeck College, Malet Street, London WC1E 7HX.
}    

\maketitle

\begin{abstract}

The proposal that the one-parameter solutions of the real part of the Schr\"{o}dinger 
equation (quantum Hamilton-Jacobi equation) can be regarded as `quantum particle
trajectories' has received considerable attention recently.  Opinions as to their
significance differ.  Some argue that they do play a fundamental role as actual
particle trajectories, others regard them as mere metaphysical appendages without any
physical significance.  Recent work has claimed that in some cases the Bohm approach
gives results that disagree with those obtained from standard quantum mechanics and,
in consequence, with experiment. Furthermore it is claimed that these trajectories have
such unacceptable properties that they can only be considered as `surreal'.   We
re-examine these questions and show that the specific objections raised by Englert,
Scully, S\"{u}ssmann and Walther cannot be sustained. We also argue that contrary to
their negative view, these trajectories can provide a deeper insight into quantum
processes.
\end{abstract}


\section{Introduction}\label{sec.1}

The significance of the one-parameter solutions of the modified Hamilton-Jacobi 
equation of Bohm \cite {Bohm52} has been the  subject of many discussions over the years.
(For more details of this approach see Bohm and Hiley \cite{BH87} \cite{BH93}, Holland
\cite{PH}  and D\"{u}rr, Goldstein
and Zanghi \cite{DGZ})  Attempts to explore these solutions and to give them physical
significance in terms of particle trajectories has often been met with strong
opposition.  For some they are merely `metaphysical baggage' with no real physical
significance and should therefore not be pursued further 
(Pauli \cite{PAU} and Zeh \cite{Z}). 
Relatively recently Englert, Scully, S\"{u}ssmann and Walther [ESSW2] \cite{ESSW2}  and
Scully
\cite{S} have claimed to show that these `trajectories' lead to results that disagree
with the standard interpretation of quantum mechanics.  However their `standard' interpretation is not the
one that is usually called the Copenhagen interpretation.  Furthermore they claim that
these `trajectories' have such bizarre properties that they cannot possibly be
considered as `real' particle trajectories and must be regarded as `surreal', thus
having no physical significance. 

The specific conclusions of ESSW2 were answered by Dewdney
et al \cite{DHS} and by
D\"{u}rr et al \cite{DF}, who use the term Bohmian Mechanics to describe their own
version of what was introduced by Bohm \cite{Bohm52} and developed by Bohm and
Hiley \cite{BH87} \cite{BH93}.  However their answers have not removed the perceived
difficulties as was shown in Englert et al [ESSW3]
\cite{ESSW3}  and in a more recent paper Scully \cite{S} 
 repeats the criticisms contained in the earlier papers of ESSW2 and ESSW3. 

In
ESSW3 we find the statement that 
\begin{quote}
Nowhere did we claim that BM makes predictions that
differ from those of standard quantum mechanics.
\end{quote}
  Yet in ESSW2 we find
\begin{quote}
 In other words: the Bohm
trajectory is here macroscopically at variance with the actual, that is: {\em observed}, track.
\end{quote}
 Again in
Scully \cite{S} we find 
\begin{quote}
These
[Bohm] trajectories are not the ones we would expect from QM which predicts that atoms
go along path 2 into detector A (our D$_{2}$) and along 1 into B (our 
D$_{1}$). (See figure 1
below).
\end{quote}
  Either the predictions are the same, in which case there is no reason to
favour one approach over the other except for personal preferences, or the predictions
are different in which case we can allow experiment to decide. We will 
see that no experiment can decide between the standard interpretation 
and the Bohm interpretation.  The conclusions to the contrary that were 
claimed by ESSW2 involve adding an additional assumption to the standard 
theory that leads to an internal contradiction in their work.   

The alleged difference arises
from a consideration of the experiment shown in figure 1.  If a measurement of the
energy of the cavity after the particle is detected at D$_{2}$ is found to be in an excited
state, ESSW2 conclude that ``the atom must have actually gone through the cavity''.  It
is easy to show that not all the Bohm trajectories actually go through the cavity even
though it is left in an excited state (see figure 5). Thus the two approaches appear
to lead to contradictory results.  

There is no disagreement about how the Bohm trajectories behave. The disagreement arises from the answer
to question: ``Is it correct, within
orthodox quantum mechanics, to conclude that an atom must have actually gone through the cavity when we
find it in an excited state after the atom has been detected in D$_{2}$?''  We will discuss this question
in detail in section 2 of this paper, but it should be noted that in order to reach this conclusion we must
assume that it is {\em only} when a particle  {\em actually} goes through
such a detector that an exchange in energy is possible. 
This is an additional assumption that is not part of the interpretation 
proposed by Bohr and the Copenhagen school
and is {\em not} part of standard quantum mechanics. 

 We will show that this assumption leads to the well-known
contradiction that when interference effects are  involved, we
are obliged to say that, on the one hand, the atom always chooses one of the two ways,
but behaves as if it had passed both ways \cite{NB61}.  If our objections are correct then
the Bohm trajectories cannot be ruled out by the arguments presented in ESSW2.

Further objections to the
Bohm approach have been made by Aharonov and Vaidman \cite{AV} and by 
Griffiths \cite{RG}{\footnote {We will discuss this paper elsewhere.}}. 
Unfortunately, in the case of Aharonov and Vaidman \cite{AV}, they have not used the
approach introduced by Bohm \cite{Bohm52} and further developed in Bohm 
and Hiley \cite{BH87} \cite{BH93}. In
their own words ``The fact that we see these difficulties follow from 
{\em our [AV] particular
approach} to the Bohm theory in which the wave is not considered to be a `reality'.''
\cite{AV} 

The basic assumptions
used in Bohm and Hiley \cite{BH93} are set out in their book, {\em The Undivided Universe}.  Assumption
1 defines the role played by the particle and is the same as assumption 1 in Aharonov and Vaidman, but
assumption 2 reads ``This particle is never separate from a new
 type of field that
fundamentally affects it.  This field is given by $R$ and $S$ or alternatively 
by $\psi = R\mbox{
exp}(iS/\hbar)$.  $\psi$ satisfies the Schr\"{o}dinger equation (rather than, for example,
Maxwell's equation), so that it too changes continuously and is causally determined.''
Aharonov and Vaidman have replaced this last assumption by one that does not give the
wave function the same role and, as a consequence, their criticisms do not apply to the
Bohm approach we discuss in this paper.  Nevertheless they have raised an interesting
question concerning tracks produced in a bubble chamber, which we will address in
section 5.5.

Our assumption 2 is the source of a number of features of quantum
processes that, for one reason or another have been regarded as undesirable or
unacceptable, ``quantum non-locality'' or ``quantum non-separability'', being, perhaps,
the most contentious.  This feature clearly arises in our approach and was used by
Dewdney et al \cite{DHS} to explain the `strange' behaviour of the trajectories.  It seems
that this non-local or non-separable feature disturbs ESSW3 because they 
write 
\begin{quote}
It is
quite unnecessary, and indeed {\em dangerous}, to attribute any additional ``real'' meaning to
the $\psi$-function.
\end{quote}
 Unfortunately the specific `dangers' are not spelt out.   

The opposition
to non-separability is deeply entrenched in spite of all that Bohr has written about
quantum theory. He constantly emphasised that the central feature of quantum theory
lay in the ``impossibility of making a sharp separation between the behaviour of atomic
objects and the interaction with the measuring instruments, which serve to define
conditions under which the phenomena appear.''\cite{NB61B} Indeed as one of us has
pointed out already  \cite{BJH99}, Bohr's answer to the original
Einstein-Podolsky-Rosen \cite{EPR} objection depends on the `wholeness of the experimental
situation' which characterises the impossibility of making this sharp separation. In this
regard the Bohm approach actually supports Bohr's conclusions, although from a point of view
that Bohr himself thought to be impossible!  Thus we find it very strange that the reason
for rejecting the Bohm approach is central to Bohr's answer to the EPR objection and
therefore, it to must be rejected.  

Unfortunately the
confusion we find in this field is not helped by the dogmatism that has become
fashionable on both sides.  These positions arise from what appear to be deeply held
convictions as to what quantum physics {\em ought to be} rather than let experiment,
mathematics and clear logic lead the debate. (After all both standard quantum mechanics and the Bohm
approach claim to use exactly the same mathematics and to predict exactly the same
experimental verifiable probabilities.)  This dogmatism has generated much confusion over the
role of `particle' trajectories in quantum mechanics.

In this paper
we will review the general situation and attempt to clarify how and where the
disagreements arise. We hope that this discussion will be received in the spirit that
it is written, namely it is an attempt to reach common ground in which both sides can
actually agree on what are the essential differences. It is important to find out
whether they are factual and amenable to experimental clarification or whether they
are merely disagreements about what satisfies our demands for ``common sense''
theories.
\section{ Appraisal of arguments presented by Englert, Scully, S\"{u}ssmann, Walther.
}\label{sec.2}
\subsection{The general grounds.}\label{subsec.2.1 }
Let us begin by considering some comments made in the recent paper by Scully 
\cite{S}.  
He asks the specific question 
\begin{quote}
Do Bohm trajectories always provide a trustworthy
physical picture of particle motion?
\end{quote}
 and immediately provides the 
answer 
``No",
followed by the explanation 
\begin{quote}
When particles detectors are included particles do not
follow the Bohm trajectories as we would expect from a classical type model.
\end{quote} 
Unfortunately this critical sentence is not very clear.  Is it saying that we somehow
know which trajectories a particle would follow in the interferometer in question, or
is it simply saying that the trajectories are not doing what we would expect from the
point of view of classical physics?  

If it is the latter, it is surely clear by now
that we cannot explain quantum processes using classical physics, so why would we
expect classical-type trajectories to account for quantum processes?  If this is what
is meant then the objection is not serious and can be dismissed immediately. 

If,
however, it is the former, then it implies that we know, independently of the Bohm
approach, which trajectories the particles {\em actually} follow in an interferometer. Later
in the same paper we find a much clearer statement confirming this 
view, which we have quoted above but which we will
repeat  again. 
\begin{quote}
These [Bohm] trajectories are not the ones we would expect from QM
which predicts that atoms go along path 2 into detector A [our D$_{2}$] and along 1 into B
[our D$_{1}$].
\end{quote}  

But how do we know what trajectories the particles actually follow in
orthodox quantum mechanics?  Is it not an essential feature of standard quantum
mechanics that when we are discussing interference or diffraction effects, talking
about trajectories will lead to contradictions?  For example, in discussing electron
diffraction, Landau and Lifshitz \cite{LL} point out that since the interference pattern
does not correspond to the sum of patterns given by each slit (beam) separately,  
    ``It
is clear that this result can in no way be reconciled with the idea that electrons
move in paths''. 

  Again Bohr \cite{NB61A}, referring to an interferometer using photons
rather than atoms, remarks:  
\begin{quote}
    In any attempt of a pictorial representation of the
behaviour of the photon we would, thus, meet with the difficulty: to be obliged to
say, on the one hand, that the photon always chooses one of the two ways and, on the
other hand, [when the beams overlap] that it behaves as if it had passed both ways. 
{\em It is just arguments of this kind which recall the impossibility of subdividing
quantum phenomena and reveal the ambiguity in ascribing customary physical attributes
to atomic objects}. 
\end{quote}
  He goes on to say that 
  \begin{quote}
       all unambiguous use of space-time
concepts in the description of atomic phenomena is confined to the 
{\em recording of
observations} which refer to marks on a photographic plate or to similar practically
{\em irreversible amplification} effects like building of a drop of water around an ion in a
cloud-chamber.
  \end{quote}
  On what grounds then is the claim that, in standard quantum mechanics, we can `know'
the path a particle takes as it passes through an interferometer being made?  It is
essential to get a clear answer to this question because without it, the rejection of
Bohm trajectories on the grounds that it disagrees with quantum mechanics cannot be
sustained.  How then are trajectories to be determined in standard quantum
mechanics?   

The experiment central to this discussion is the interferometer sketched
in figure 1
\begin{figure}[t]
\includegraphics{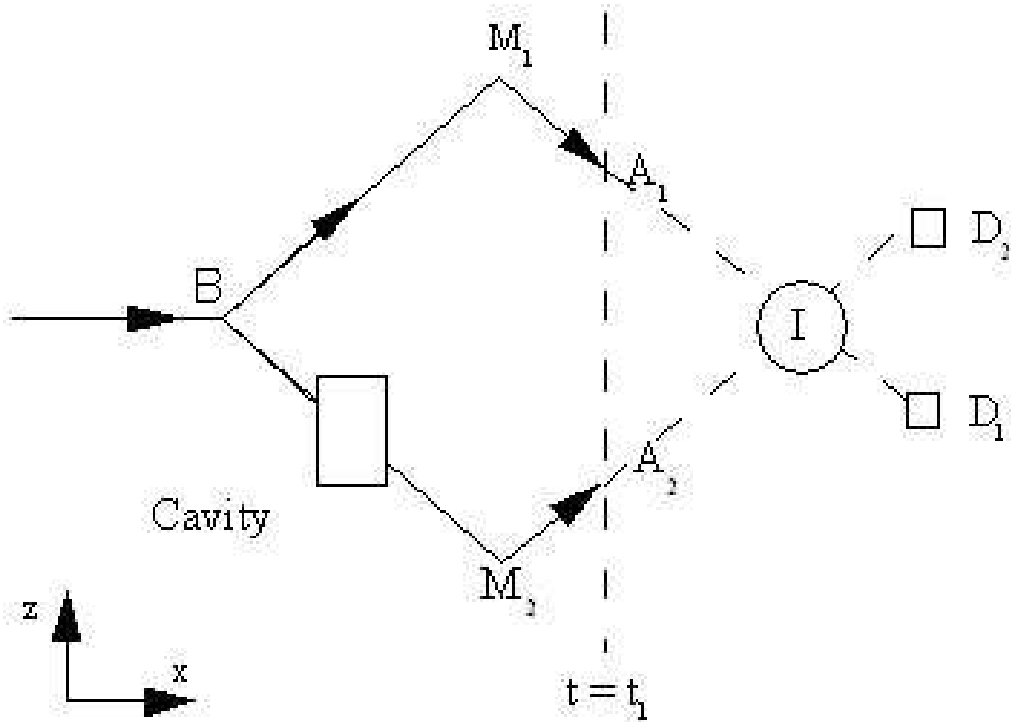}
\caption{The interferometer considered by Scully \cite{S}.}
\end{figure}

A beam of atoms is incident on a beam-splitter B.   Each atom is assumed to be in an 
excited (Rydberg) state. It is further assumed that each atom can be represented by a
Gaussian wave packet of small width so that after passing through the beam-splitter,
B, the wave packets follow the two paths $BM_{1}A_{1}$ and $BM_{2}A_{2}$ and do not overlap again until
they reach the region
$I$. 

Before reaching this
region, a special cavity micromaser is placed in one of the arms of the interferometer
as shown in figure 1. The aim of introducing this cavity is to provide 
a {\em Welcher Weg}
(WW) device, which we can use to enable us to {\it infer} the path the atom took in passing
through the interferometer.   

The cavity has the essential property that when an
excited atom is passed through it, the excitation energy is exchanged with the cavity,
and the atom then continues in the same direction with the same momentum, but in a
lower internal energy state.  This means that when the cavity is part of an
interferometer as shown in figure 1, the exchange of energy does not destroy the
coherence between the two beams. 

It is then claimed that, by measuring the energy in the cavity
after the atom has passed through the interference region $I$, we will be able to {\it infer}
along which arm the atom actually went.  If this claim is correct then not only does it
rule out the Bohm approach, it also throws doubts on the validity of the assertions
made by Landau and Lifshitz and by Bohr in the quotations presented above.  The
crucial question then is whether the cavity used in this way can give reliable
information as to which path the atom actually took. 

To answer this, it is crucial to
analyse carefully how the cavity and the atom exchange energy, as it is this exchange
process that lies at the heart of the objections raised by ESSW2 and Scully 
\cite{S}. 
Our analysis will involve taking a careful look at what assumptions these authors make
when they refer to `the standard approach to quantum mechanics.'  

One primary
assumption made in ESSW3 is that the wave function is merely a ``tool used by
theoreticians to arrive at probabilistic predictions.''  This means that we must first
analyse the interaction between the cavity and the atom {\em classically} so that a suitable
interaction Hamiltonian can be written down. This Hamiltonian is then written in an
{\em operator form} so that we can use the Schr\"{o}dinger equation to solve for the time
development of the wave function.  Thus as far as quantum formalism is concerned, the
interaction involves a change of relationship between the {\em wave function} of the atom
and the {\em wave function} of the cavity.  The quantum formalism does not require any knowledge of the
position of the atom.  The formalism enables us to calculate the probability of finding an atom at any
given point at any given time.

Since the interaction
Hamiltonian is local and the atom is represented by a Gaussian wave packet of narrow
width, the corresponding ket for the whole system at time $t = t_{1}$ (See figure 1) is 

\begin{equation}
    |\Psi(t=t_{1})\rangle=|\psi_{1}\rangle|\Phi_{0}\rangle+|\psi_{2}\rangle|\Phi_{E}\rangle
    \label{eq:1}
\end{equation}
where $|\psi_{1}\rangle (|\psi_{2}\rangle)$ is the ket of the excited (unexcited) atom
and $|\Phi_{0}\rangle (|\Phi_{E}\rangle)$ is the ket of the
 unexcited (excited) cavity. In writing down this expression we are assuming that no
irreversible process takes place when the cavity is added.\footnote {The de-excited atom that
has passed through the cavity can be reflected back into the cavity, so that it can become excited
again..}  
In terms of wave packets, we can write
$\psi_{1}(\mathbf{r}_{a})\phi_{1}(\mathbf{\eta}_{a})=\langle\mathbf{r}_{a}, 
\mathbf{\eta}_{a}|\psi_{1}\rangle$   and 
$\psi_{2}(\mathbf{r}_{a})\phi_{2}(\mathbf{\eta}_{a})=\langle\mathbf{r}_{a},\mathbf{\eta}_{a}|\psi_{2}\rangle$
where $\psi$ is the wave function of the centre of mass of the atom 
with centre of mass 
co-ordinates  $\mathbf{r}_{a}$  and $\phi$ in the internal wave 
function depending on the variables $\mathbf{\eta}_{a}$. 

After the wave packets have  separated, they do not overlap prior to
$t = t_{1}$, so we can talk about `the wave packet $\psi_{1}(\mathbf{r}_{a}$) travelling along the path
BM$_{1}$A$_{1}$' and `the wave packet $\psi_{2}(\mathbf{r}_{a}$)
travelling along BM$_{2}$A$_{2}$'. At this stage we can associate an atom with a particular wave
packet
and talk about the `atom travelling along BM$_{1}A_{1}$' or `along 
BM$_{2}$A$_{2}$' without running into
any difficulties.  

However once the packets overlap again as they enter the region $I$, we
must proceed with caution, particularly in view of the warnings given by 
Bohr \cite{NB61}
and Landau and Lifshitz \cite{LL} in the quotations above.  The key objection they raise
is that if we give relevance to the atom as opposed to the wave function, we are
forced to say that the atom always chooses one path, but behaves as if it had passed
both ways.  How has this objection been avoided in ESSW2?

Let us concentrate on the
region of overlap $I$.  We have argued that the interaction with the cavity does not
destroy the coherence between the two beams when they subsequently overlap again. 
This means that the beam can be legitimately described by the ket 
$|\Psi\rangle$   given by equation
(1).  However if the interaction were to destroy the coherence, then we must replace
this ket by the density operator
\begin{equation}
      \rho=|\psi_{1}\rangle|\Phi_{0}\rangle\langle\psi_{1}|\langle\Phi_{0}|+|\psi_{2}\rangle|
    \Phi_{E}\rangle\langle\psi_{2}|\langle\Phi_{E}|
    \label{eq:2}
\end{equation}
This would describe two  separate wave packets moving through the region $I$ without
 interfering with each other.  In other words when an atom enters the beam 
 BM$_{1}$A$_{1}$, it
is confined to the packet $\psi_{1}$ which then moves along the path 
BM$_{1}$A$_{1}$ and continues
without deviation through the region $I$, finally arriving at D$_{1}$. An atom following
the path BM$_{2}$A$_{2}$ will be in the wave packet $\psi_{2}$ and it too will continue also without
deviation through the region $I$ until it is registered at D$_{2}$.  

In this case there is no
doubt that the atom actually follows one or other of the paths and that the atom that
went through the cavity exchanged energy with it and then continued on 
to D$_{2}$.  There is no problem here\footnote {It should be noted that in this case  
the Bohm theory will also produce trajectories that cross in the region {\em I} so that 
in this case there is 
no disagreement (see section 5.4).}


\subsection{The region of coherent overlap.}\label{subsec.2.2}
The case that does present difficulties is the one that arises only when coherence is
maintained.  Here the correct description of the experimental set-up is provided by
the ket $|\Psi\rangle$.  Let us consider the situation at a time $t = 
t_{3}$, after the atom has passed
through the region $I$ so that there is no longer any overlap between the two wave
packets. The experimental predictions are quite clear.  If D$_{1}$ fires, we will find the
cavity is unexcited, whereas if D$_{2}$ fires, the cavity will be found to be excited.

This conclusion is reached in {\it both} interpretations.  No assumptions about
possible particle trajectories are needed to arrive at this conclusion.  Clearly this result
is quite consistent with the statement `the atom passed through the cavity on its way
detector D$_{2}$'.  However {\em consistency} does not mean that  the atom did {\em
actually} go through the cavity. 

The key question then is ``How can we show which way the atom reaching D$_{1}$ or D$_{2}$
{\em actually} went in either case?''  ESSW3 claim that we can do this within the framework of
standard quantum mechanics and it is the presence of the cavity that enables us to
talk about ``the {\em detected, actual} way through the interferometer.''  This is the key
statement upon which ESSW base all their conclusions and it must be examined very
carefully. 

The first point to notice is that the cavity, which ESSW are regarding 
as a `measuring' device, does not
function in the same way as a traditional measuring device in standard quantum mechanics.  The cavity is a
quantum system and no irreversible mark has been left in any system.  Since it leaves no
irreversible mark, it is not a measuring device in the traditional sense as defined by
Bohr (See  quotation above). ESSW claim
that the cavity gives us  a new type of measurement, which does not leave a permanent 
record and can be easily
`erased'\cite{SEW}.  Thus ESSW are talking about an aspect of quantum mechanics that is not contained
in the Copenhagen interpretation.

In order to emphasise the difference, let us look more carefully at orthodox measurement.  To
discuss a measurement Bohr  introduces the word `phenomenon' defined in the following way: 
\begin{quote}
 As a more appropriate way of expression I advocate the application of the word {\em phenomenon}
exclusively to refer to the observations obtained under specified circumstances, including an account of
the whole experimental arrangement \cite{NB64}. 
\end{quote} 
But we must take it further.  As Wheeler
\cite{JW} puts it: ``No elementary quantum phenomenon is a phenomenon until, in the words of Bohr
\cite{NBPR} `It has been brought to a close' by `an irreversible act of amplification'".  In
drawing attention to this traditional thinking we are not, at this stage making any value
judgements.  We are simply drawing attention to the fact that ESSW have added something new
to what we would  call `standard quantum mechanics'.

The key assumption used by ESSW is that energy exchange only takes place when the actual
atom interacts {\em locally} with the photon field in the cavity.  This does not merely mean
that the interaction Hamiltonian is local, but that the interaction can take place if
and only if {\em the atom is physically present} in the cavity.  {\em This is a new assumption
that is not part of standard quantum mechanics and certainly not part the Bohr (Copenhagen)
interpretation}. 

Given this new assumption, the question that we must examine is: ``Can we
give a consistent account of quantum interference phenomena without running into the difficulties pointed
out by Bohr and by Landau and Lifshitz above?" 

        As we will be interested in the region of interference $I$ in figure 1, let us insert a horizontal 
beam-splitter at $I$.  This turns the experimental set-up into a Mach-Zender 
interferometer, which will enable us to demonstrate unambiguously the interference properties that 
occur in region $I$.  

Without the
cavity in the arm BM$_{2}$A$_{2}$, we find for the symmetrical set-up that all the atoms end up in D$_{2}$.
If we now include the cavity in the arm BM$_{2}$A$_{2}$, we find the probability for D$_{1}$  firing is
 given by the expression
\begin{eqnarray}
P(D_{1}) = 1/4[ \langle \psi_{1}|\psi_{1}\rangle \langle\Phi_{0}|\Phi_{0}\rangle + 
\langle \psi_{2}|\psi_{2}\rangle \langle\Phi_{E}|\Phi_{E}\rangle \nonumber \\+ \langle
\psi_{1}|\psi_{2}\rangle
\langle\Phi_{0}|\Phi_{E}\rangle  + \langle \psi_{2}|\psi_{1}\rangle \langle\Phi_{E}|\Phi_{0}\rangle]
\end{eqnarray}
While the probability for $D_{2}$  firing is
\begin{eqnarray}
P(D_{2}) = 1/4[\langle \psi_{1}|\psi_{1}\rangle
\langle\Phi_{0}|\Phi_{0}\rangle + 
\langle\psi_{2}|\psi_{2}\rangle \langle\Phi_{E}|\Phi_{E}\rangle \nonumber \\ - \langle
\psi_{1}|\psi_{2}\rangle\langle\Phi_{0}|\Phi_{E}\rangle - \langle \psi_{2}|\psi_{1}\rangle
\langle\Phi_{E}|\Phi_{0}\rangle]
\end{eqnarray}
We see that we get a new firing probability for the detectors depending on whether
(a) $|\psi_{1}\rangle$  is orthogonal to $ |\psi_{2}\rangle$,
(b) $|\Phi_{0}\rangle$  is orthogonal to $|\Phi_{E}\rangle$
or (c) both (a) and (b).

All of this is very obvious and straight forward, but now the assumption that ESSW make is that 
the atom must have actually gone through the detector to exchange energy.  ESSW3 write:
\begin{quote}
-the interpretation of the Bohm trajectory -- is implausible, because this trajectory can be
macroscopically at variance with the detected, actual way through the interferometer.  And yes, we do
have a framework to talk about path detection; it is based upon the local interaction of the atom with the
photons inside the resonator, described by standard quantum theory with its short range interactions only.
\end{quote} 
Thus the claim is that the {\em  only} way that the cavity can be excited 
is if the atom actually passes through it. 
 This means that with the cavity in place 50\% of the atoms actually go 
 through the cavity and end up triggering D$_{2}$.¥
 The remaining
50\% actually pass down the other arm and end up triggering D$_{1}$.  

However when the cavity is removed, {\em all} the particles end up in $D_{2}$.  How then does one explain
 why the particles travelling down BM$_{1}$A$_{1}$ stop travelling on to $D_{1}$  and instead travel
 to
$D_{2}$?  Nothing has been changed in path $1$ yet somehow the particles travelling along 
path $1$  `know' the cavity is present in path $2$ or not as
the case may be?  

Recall that ESSW insist that only short range interactions are allowed in
standard quantum mechanics.  There is no explanation of this change of 
behaviour and we are simply left with the contradiction that
Bohr\cite{NB61A}, and Landau and Lifshitz\cite{LL} have already pointed 
out, namely, that ``the photon
always choose one of two ways" but ``behaves as if it had passed both ways." 

The above results indicate that coherence is maintained and the absence of interference should not be taken to mean a loss of coherence between the two beams
 in the region {\em I}. These two beams must be treated as remaining coherent.  This would
 certainly not
be the case if we measured the energy in the cavity {\em before} the atom reached the region {\em I}.  But
that would require an irreversible process to occur in the recording of the result.  In that case, in
Wheeler's terms ``the phenomenon is complete" and we have  information that will enable us to say
along which path the atom actually travelled.  Mathematically this would mean replacing the
wave function (1) by the density operator (2).

It might be argued that it is the exchange of energy that is responsible 
for the lack of interference or `decoherence'.  However 
this cannot be true.  If we add any device that interacts with the atom, energy must be exchanged
even if
this energy induces only a change of phase.  This would occur if the 
atom were to interact with an oscillator in a
coherent state.  Since coherent states are not orthogonal, the interference does not disappear, showing
that merely an exchange of energy is NOT responsible for decoherence.

\subsection{How the Copenhagen interpretation deals with this situation.}
\label{subsec.2.3 }
The traditional way to avoid all of these difficulties is to give up any
attempt to follow a particle along a well-defined path, particularly 
in an interferometer. This does not mean
that we can never talk about the path of a particle.   
Heisenberg \cite{WH} pointed
out in 1927 that before we can talk about a path, we have to be clear as to what is to be
understood by the words ``position of the object''.   He writes 
\begin{quotation}
    When one wants to be
clear about what is to be understood by the words ``position of the object'', for
example of the electron, then one must specify experiments with which whose help one
plans to measure the ``position of the electron''; otherwise this word has no meaning.
\end{quotation}

Conventional measurement requires the observable to be
represented by an operator and after the measurement is complete, the particle is left
in an eigenstate of the operator.  ESSW2 specifically rule out any change of the
centre-of-mass motion and therefore the atom is not in a position eigenstate after it
leaves the cavity. Nor is it a `detection' in the same sense as when the atom is
recorded at D$_{1}$ or D$_{2}$.  Here some form of irreversible amplification involved.  Rather
their notion of measurement involves {\em inference} based on the assumption that energy
exchange can {\em only} take place when the atom is physically present in the cavity. 

There is no difficulty here if the energy of the cavity is measured 
before the atom reaches $I$.  However if we leave this measurement 
until after the atom has passed through $I$, the wave functions 
obtained from equation (1) shows that there is a coupling between 
$\psi_{1}(\mathbf{r}_{a})$ and $\psi_{2}(\mathbf{r}_{a})$.  Bohr and Wheeler argue that 
this coupling cannot be ignored until the whole process is ``brought 
to a close by an irreversible act of amplification''.  If we do ignore 
the coupling and follow ESSW we are led to the contradiction 
that ``although the
particle travels down one path of an interferometer, it behaves as if it went down
both paths.''

This is just what the Copenhagen interpretation warns us about.  To
emphasise this point again, consider the following quotation taken from 
Bohr \cite{NB51}:
\begin{quotation}
    In particular, it must be realised that - besides in the account of the placing
and timing on the instruments forming the experimental arrangement - all unambiguous
use of space-time concepts in the description of atomic phenomena is the recording of
observations which refer to marks on a photographic or similar practically
irreversible amplification effects like the building of a water drop around an ion in
a cloud-chamber.
\end{quotation}
  Thus the storage of a single quantum of energy in the cavity does not constitute a
measurement.  There is no `irreversible
amplification' until the atom is detected in D$_{1}$ or D$_{2}$.  

As we have already remarked above, if we measure the energy in
the cavity {\em after} the atom passes through the cavity but {\em before} it reaches the region
$I$, an irreversible change does take place and the coherence between the two beams is
subsequently destroyed. In this case we must use the density operator (2) in the region
$I$  and then we can unambiguously conclude that the atom
passed through the cavity and its energy can be used to infer that the atom passed
through the cavity.  But once we allow the beams to intersect in the 
region $I$, we can
no longer make this inference without making the assumption that leads to the
contradiction discussed above. This is why Heisenberg \cite{WH58} wrote: 
\begin{quotation}
    If we want to
describe what happens in an atomic event, we have to realise that the word `happens'
can only apply to observations, not to the state of affairs between two observations.
\end{quotation}
The notion of a WW device has no meaning whatsoever in the Copenhagen interpretation
and we cannot use it to give a meaning to which way the particle passed through the
interferometer. The inference that the energy in the cavity can reveal 
what path the atom took is incorrect once the atom has entered the 
region $I$ and as a consequence the claim that one can use quantum mechanics to show
that the Bohm trajectories are ``meaningless'' cannot be sustained.


\section{Probability currents}\label{sec.3}
One of the features of the Bohm theory that generates misgivings is the fact that it 
predicts that atoms do not cross the $z = 0$ plane of symmetry  when there is no
cavity in either arm (see figure 3).  This result has been greeted
with surprise, if not disbelief.  How is it possible for atoms to be so drastically
deflected when there appears to be nothing in the region $I$ that could bring about
this reflection? Before answering this question from the Bohm perspective, let us
first see if there is anything within the orthodox interpretation that might enable us
to find some way of exploring what might be going on the region $I$. 

Central to standard quantum mechanics is probability and its conservation as
expressed through the equation
\begin{equation}
\frac{\partial P}{\partial t} + \nabla.{\mathbf{j}} = 0
\end{equation}
In order to conserve probability as a {\em local} probability density we need to interpret
$\mathbf{j}$ as probability current. It is in this way that 
orthodox quantum mechanics allows us to talk meaningfully about probability currents.
 
Indeed  these currents are used  extensively in many branches of quantum physics including 
scattering theory, condensed matter physics and superconductivity, where we can discuss
the flow of charge across boundaries.  We can interpret these currents
without changing the significance of the wave function, which can still be regarded as a
`tool', forming part of an algorithm.  It is this algorithm that enables us calculate not only
probabilities, but also probability currents in any given experimental set-up.   Could these
probability currents supply further information about the flux of atoms in the region of
interference,
$I$, and hence about the flux across the $z = 0$ plane?  

It is interesting to note that ESSW2
actually use these currents to criticise the Bohm interpretation, attributing their
properties to the Bohm interpretation and do not seem to realise that the probability
currents are part of the orthodox quantum mechanics\footnote {As we will show below,
the Bohm approach uses  equations that have exactly the same mathematical form as these
currents, but because the  meaning of the wave function is changed and the particle is
given a well-defined role, the conclusions drawn from these equations are different. 
However the conclusions drawn from  the probability currents do not contradict those
arising from the Bohm approach.}

        Let us first use the probability currents to explore the difference between the 
situation described by the ket $|\Psi\rangle$   given by equation (1) and the density 
operator $\rho$
given by equation (2).

        The probability current is defined by
\begin{equation}
    \mathbf{j}=\frac{\hbar^{2}}{2mi}[\Psi^{*}(\nabla\Psi)-(\nabla\Psi^{*})\Psi]
    \label{eq:6}
\end{equation}
In the case of the density operator each term gives rise to an independent current, 
$\mathbf{j}_{1}$ and $\mathbf{j}_{2}$, where
\begin{equation}   
\mathbf{j}_{i}=\frac{\hbar^{2}}{2mi}[\Psi_{i}^{*}(\nabla\Psi_{i})-(\nabla\Psi_{i}^{*})\Psi_{i}¥]
    \hspace{1cm}(i=1,2)
    \label{eq:7}
\end{equation}
These latter currents cross the region $I$ from A$_{1}$ to D$_{1}$ and 
from A$_{2}$ to D$_{2}$ respectively.
  In other words the currents do not `see' each other and there is no ambiguity as to
what is happening in this case. Indeed the consideration of the probability currents
merely confirms our previous discussion.

Let us now turn to consider what happens in
the situation described by the ket $|\Psi\rangle$. We first consider the case when no cavity is
present in either arm of the interferometer.  Here the wave function is simply
\begin{equation}
    \Psi_{I}(\mathbf{r}_{a})=\psi_{1}(\mathbf{r}_{a})+\psi_{2}(\mathbf{r}_{a})
    \label{eq:8}
\end{equation}
The probability current for the atoms in the region of overlap $I$, is given by
\begin{eqnarray}
\mathbf{j}(\mathbf{r}_{a})=
\frac{\hbar^{2}}{2mi}[\psi_{1}^{*}(\mathbf{r}_{a})\nabla\psi_{1}(\mathbf{r}_{a})-
    \psi_{1}(\mathbf{r}_{a})\nabla\psi_{1}^{*}(\mathbf{r}_{a})]
    +[\psi_{2}^{*}(\mathbf{r}_{a})\nabla\psi_{2}(\mathbf{r}_{a})-
    \psi_{2}(\mathbf{r}_{a})\nabla\psi_{2}^{*}(\mathbf{r}_{a})]\nonumber 
    \\
    +[\psi_{2}^{*}(\mathbf{r}_{a})\nabla\psi_{1}(\mathbf{r}_{a})-
    \psi_{1}(\mathbf{r}_{a})\nabla\psi_{2}^{*}(\mathbf{r}_{a})]+
    [\psi_{1}^{*}(\mathbf{r}_{a})\nabla\psi_{2}(\mathbf{r}_{a})-
    \psi_{2}(\mathbf{r}_{a})\nabla\psi_{1}^{*}(\mathbf{r}_{a})]
\end{eqnarray}
The first two terms in this expression are exactly the currents $\mathbf{j}_{1}$ and
$\mathbf{j}_{2}$ calculated
 using the density operator
 \begin{equation}
     \rho=|\psi_{1}\rangle\langle\psi_{1}|+|\psi_{2}\rangle\langle\psi_{2}|
     \label{eq:10}
 \end{equation}
The third and forth terms correspond to the interference terms.

        We could deal with this numerically and calculate the probability current in detail,
 but our main point can be made using the same argument employed by ESSW2.   
 This depends on the
fact that $\psi_{1}(\mathbf{r}_{a})$ is the reflected image of $\psi_{2}(\mathbf{r}_{a})$ about 
the horizontal line of
symmetry ($z = 0$) in figure 1.\footnote {This fact was also used in the attempt to
discredit the  Bohm approach.} 
This means that
\begin{equation}
    \psi_{1}(x,y,z,t)=\psi_{2}(x,y,-z,t)
    \label{eq:11}
\end{equation}
In consequence the $x$- and $y$-components of the current vector is an even function 
in $z$, while the 
$z$-component is odd.  Therefore the $z$-component of the current is an odd function of $z$,
so that $j_{z} =0$ at $z = 0$.  Thus there is no probability current flowing across the
horizontal plane, and hence no net particle-flux across this plane.

        This could imply either (a) that no particles actually cross the $z = 0$ plane,
or (b) 
that the average particle flux crossing the plane is zero.  But (b) is exactly the
same as the result of using the density operator (9) when we simply add the two
independent currents together. Clearly in this case there are as many particles travelling
in the negative $z$-direction as there are crossing in the positive $z$-direction. 

This could well be the case when
we use the wave function (7) but now we cannot split the ensemble into two separate
sub-ensembles because of the presence of observable interference effects.  Therefore we cannot
be sure that zero current arises because there are as many atoms crossing the $z = 0$ plane one
way as the other.
Thus at best we are left with an ambiguity of being unable to decide between the
two choices (a) and (b), but we certainly cannot rule out possibility (a) within quantum
mechanics.

In order to explore the nature of this ambiguity further, let us consider a more general case
when an energy-exchanging device, d$_{e-e}$ is placed in one of the
arms. We will again assume this device is a single state microscopic quantum system of
some kind that does not introduce any irreversible effects. This could be, for
example, a harmonic oscillator in an energy eigenstate, an oscillator in a coherent
state, or some other form of phase shifter. It could even be an idealised system such
as `particle in an infinite well' (i.e. a particle in a `box'.) But no matter what the
device is, we assume that some energy will be exchanged between the device and the
atom.

 Consider the case when the atom leaves the device d$_{e-e}$ in an excited
 state that is {\em not} orthogonal to its ground state (for example, in a
harmonic oscillator in a coherent state).  Let the normalised wave function of this
device before the interaction be $\eta_{0}(\mathbf{r}_{b}$) and after the interaction
be $\eta_{1}(\mathbf{r}_{b}$).  Here $\mathbf{r}_{b}$ is the position 
of the particle comprising the harmonic oscillator. Assuming
coherence is not destroyed in the region {\em I}, the wave function at $t > t_{1}$ will be
\begin{equation}       
      \Psi_{1}(\mathbf{r}_{a},\mathbf{r}_{b})=\psi_{1}(\mathbf{r}_{a})\eta_{0}(\mathbf{r}_{b})+
    \psi_{2}(\mathbf{r}_{a})\eta_{1}(\mathbf{r}_{b})
\end{equation}
We can form $|\Psi_{1}(\mathbf{r}_{a},\mathbf{r}_{b})|^{2}$ and integrate 
over $\mathbf{r}_{b}$ to obtain an expression for the 
probability of finding an atom at a particular point $\mathbf{r}_{a}$ in the region $I$.  This is
\begin{equation}
    P(\mathbf{r}_{a})=|\psi_{1}|^{2}+|\psi_{2}|^{2}+\alpha\psi_{1}^{*}\psi_{2}+
    \alpha^{*}\psi_{1}\psi_{2}^{*}
\end{equation}
where $\alpha =\int \eta_{0}^{*}\eta_{1}d^{3}\mathbf{r}_{b}$.

        It is clear from this expression that interference effects will be seen in the region
 $I$ along any $x = constant$ plane as long as $\alpha \neq 0$ (see figure 1). Let us see what effect
this interference has on the probability currents.

In this case the conservation of probability is expressed through
\begin{equation}
\frac {\partial P(\mathbf{r}_{a},\mathbf{r}_{b})}{\partial t} +
\nabla_{a}.\mathbf{j}_{a}({\mathbf{r}_{a},\mathbf{r}_{b}}) +
\nabla_{b}.\mathbf{j}_{b}({\mathbf{r}_{a},\mathbf{r}_{b}}) = 0
\end{equation}
Here we have two currents, the first is the probability current for the atoms, which is given by
\begin{eqnarray}
\mathbf{j}_{a}(\mathbf{r}_{a},\mathbf{r}_{b})=
\frac{\hbar^{2}}{2mi}\{[\psi_{1}^{*}(\mathbf{r}_{a})\nabla_{\mathbf{r}_{a}}\psi_{1}
(\mathbf{r}_{a})-\psi_{1}(\mathbf{r}_{a})\nabla_{\mathbf{r}_{a}}\psi_{1}^{*}
(\mathbf{r}_{a})]|\eta_{0}(\mathbf{r}_{b})|^{2}\nonumber\\
    +[\psi_{2}^{*}(\mathbf{r}_{a})\nabla_{\mathbf{r}_{a}}\psi_{2}(\mathbf{r}_{a})-
    \psi_{2}(\mathbf{r}_{a})\nabla_{\mathbf{r}_{a}}\psi_{2}^{*}(\mathbf{r}_{a})]
    |\eta_{1}(\mathbf{r}_{b})|^{2}\nonumber 
    \\
    +[\psi_{2}^{*}(\mathbf{r}_{a})\nabla_{\mathbf{r}_{a}}\psi_{1}(\mathbf{r}_{a})- 
    \psi_{1}(\mathbf{r}_{a})\nabla_{\mathbf{r}_{a}}\psi_{2}^{*}(\mathbf{r}_{a})]
    \eta_{0}(\mathbf{r}_{b})\eta_{1}^{*}(\mathbf{r}_{b})\nonumber
    \\
    +[\psi_{1}^{*}(\mathbf{r}_{a})\nabla_{\mathbf{r}_{a}}\psi_{2}(\mathbf{r}_{a})-
    \psi_{2}(\mathbf{r}_{a})\nabla_{\mathbf{r}_{a}}\psi_{1}^{*}(\mathbf{r}_{a})]
    \eta_{0}^{*}(\mathbf{r}_{b})\eta_{1}(\mathbf{r}_{b})\}
\end{eqnarray}
Notice  this current is  a function of both $\mathbf{r}_{a}$ and $\mathbf{r}_{b}$, 
showing that it is in
 {\em configuration space}.  For local measurements we must find this current as a function 
 of $\mathbf{r}_{a}$ alone, and therefore we must
integrate over all $\mathbf{r}_{b}$.  If the wave functions, 
$\eta_{0}(\mathbf{r}_{b})$ and $\eta_{1}(\mathbf{r}_{b})$, are normalised, but
not orthogonal, the probability current for the atoms is
\begin{eqnarray}
    \mathbf{j}_{a}(\mathbf{r}_{a})=
\frac{\hbar^{2}}{2mi}\{[\psi_{1}^{*}(\mathbf{r}_{a})\nabla_{\mathbf{r}_{a}}\psi_{1}
(\mathbf{r}_{a})-\psi_{1}(\mathbf{r}_{a})\nabla_{\mathbf{r}_{a}}\psi_{1}^{*}
(\mathbf{r}_{a})]\nonumber\\
    +[\psi_{2}^{*}(\mathbf{r}_{a})\nabla_{\mathbf{r}_{a}}\psi_{2}(\mathbf{r}_{a})-
    \psi_{2}(\mathbf{r}_{a})\nabla_{\mathbf{r}_{a}}\psi_{2}^{*}(\mathbf{r}_{a})]
    \nonumber 
    \\
    +\alpha[\psi_{2}^{*}(\mathbf{r}_{a})\nabla_{\mathbf{r}_{a}}\psi_{1}(\mathbf{r}_{a})- 
    \psi_{1}(\mathbf{r}_{a})\nabla_{\mathbf{r}_{a}}\psi_{2}^{*}(\mathbf{r}_{a})]
    \nonumber\\
    +\alpha_{*}¥[\psi_{1}^{*}(\mathbf{r}_{a})\nabla_{\mathbf{r}_{a}}\psi_{2}(\mathbf{r}_{a})-
    \psi_{2}(\mathbf{r}_{a})\nabla_{\mathbf{r}_{a}}\psi_{1}^{*}(\mathbf{r}_{a})]
    \end{eqnarray}
Clearly if  $\alpha\neq\alpha^{*}$,  $j_{z}$ is not an odd function of 
$z$ and therefore there is now a non-zero probability current 
crossing the
$z = 0$ plane. This probability current approaches zero as either 
$\alpha \rightarrow \alpha^{*}$,or when the wave functions of the 
added device are orthogonal ($\alpha = 0$). 

What we have shown here is that by following  the standard
approach, there is no probability current crossing  the $z = 0$ plane when there is no
device in either beam.  If we include some device in which the wave functions 
$\eta_{1}(\mathbf{r}_{b})$ and $\eta_{0}(\mathbf{r}_{b})$ are not 
orthogonal, then a current must cross the $z=0$ plane. Thus interference effects produce changes
in the probability flux of the atoms.

 However when 
these two wave functions are orthogonal, as in the case of the cavity 
then again there is no net current crossing the plane.  We repeat again, although we
cannot  conclude from this that no atoms actually cross 
this plane, we cannot rule out
this possibility. In the next section we will
show that the Bohm trajectories do not cross the $z =0$ plane in this
case showing that these trajectories are certainly consistent with standard quantum 
mechanics and therefore cannot be ruled out on these grounds.

The conservation equation contains a second probability current $j_{b}({\mathbf{r}_{a},\mathbf{r}_{b}})$
This current is for the {\em device-particle} $\mathbf{r}_{b}$.  The
wave functions for this device are
$\eta_{0}(\mathbf{r}_{b})$ and
$\eta_{1}(\mathbf{r}_{b})$.  In this case the current would be
\begin{eqnarray}
    \mathbf{j}_{b}(\mathbf{r}_{a}, \mathbf{r}_{b})=
\frac{\hbar^{2}}{2mi}\{[\eta_{0}^{*}(\mathbf{r}_{b})\nabla_{\mathbf{r}_{b}}\eta_{0}
(\mathbf{r}_{b})-\eta_{0}(\mathbf{r}_{b})\nabla_{\mathbf{r}_{b}}\eta_{0}^{*}
(\mathbf{r}_{b})]|\psi_{1}(\mathbf{r}_{a})|^{2}
\nonumber\\
    +[\eta_{1}^{*}(\mathbf{r}_{b})\nabla_{\mathbf{r}_{b}}\eta_{1}(\mathbf{r}_{b})-  
\eta_{1}(\mathbf{r}_{b})\nabla_{\mathbf{r}_{b}}\eta_{1}^{*}(\mathbf{r}_{b})]|\psi_{2}(\mathbf{r}_{a})|^{2}
    \nonumber 
    \\
    +[\eta_{1}^{*}(\mathbf{r}_{b})\nabla_{\mathbf{r}_{b}}\eta_{0}(\mathbf{r}_{b})- 
    \eta_{0}(\mathbf{r}_{b})\nabla_{\mathbf{r}_{b}}\eta_{1}^{*}(\mathbf{r}_{b})]\psi_{1}(\mathbf{r}_{a})
\psi_{2}^{*}(\mathbf{r}_{a})
    \nonumber\\
+[\eta_{0}^{*}(\mathbf{r}_{b})\nabla_{\mathbf{r}_{b}}\eta_{1}(\mathbf{r}_{b})-
\eta_{1}(\mathbf{r}_{b})\nabla_{\mathbf{r}_{b}}\eta_{0}^{*}(\mathbf{r}_{b})]\psi_{1}^{*}(\mathbf{r}_{a})
\psi_{2}(\mathbf{r}_{a})
    \end{eqnarray}

Thus we see that  in conventional quantum mechanics there is a non-zero probability 
current appearing for the device-particle.  This current is different depending upon where the wave
packets are in the apparatus.  Before they reach the region of overlap {\em I}, the current consists of
only the first two terms in equation (15).  Once the atom reaches the region {\em I}, all four terms are
present. Thus the expression for the current corresponding to the {\em device-particle changes
when the atoms reach the region  I even though this region is some distance from the
device}.  Why should this happen in standard quantum mechanics if, as ESSW insist, only
short range forces appear in standard quantum mechanics?  

Furthermore since standard quantum mechanics actually predicts the possibility of a
current 
$\mathbf{j}_{b}$, how can we be sure that by measuring the energy of the cavity {\it
after} the atom has been detected in, say, $D_{2}$ the cavity did not change its
quantum state?  Changes in probability currents imply changes in probabilities.  Since
standard quantum mechanics implies there is a possible change  of probability
of finding  the cavity in a given quantum state as the atom passes through the region
$I$, how can we be sure that  the cavity in an excited state {\it necessarily} implies
that the atom must have passed through the cavity
$I$?   

Let us examine if there are any observable consequences of such a change.  First we
notice that if the states for the particle in a box,  
$\eta_{0}(\mathbf{r}_{b})$ and $\eta_{1}(\mathbf{r}_{b})$, are stationary states, then
$\mathbf{j}_{b}$ is always zero and we will not detect any change at all.  

If these states are not stationary then
we have non-zero currents and therefore we may have the possibility of recording the change in the
value of this current as the atoms reach {\em I}. 

The only way to observe such a change is to  measure $\mathbf{j}_{b}(\mathbf{r}_{a},
\mathbf{r}_{b})$ directly.  We do not see how to do this in practice but it clearly
depends on measuring some suitable correlations between an atom in the region {\em I}
and the device-particle.  It must be noted that this current is calculated directly
from the wave function (1) which gives rise to Einstein-Podolsky-Rosen and other
Bell-inequality violating correlations.  These correlations have been observed for a
number of different  physical situations and they  are in complete agreement with
standard quantum mechanics. 

 However if we are  measuring the current at $\mathbf{r}_{b}$ only,  we must average over all
$\mathbf{r}_{a}$ to find an expression for the current in terms of the
$\mathbf{r}_{b}$ alone.  In the Gaussian wave packets considered above, it is easy to
show that the integral over $\psi_{1}(\mathbf{r}_{a})\psi_{2}^{*}(\mathbf{r}_{a})$
and  $\psi_{1}^{*}(\mathbf{r}_{a})
\psi_{2}(\mathbf{r}_{a})$ is negligible.  So once again we see no consequences of the appearance
of this second term.  This is what we would expect as any different result would violate the no
nonlocal signalling theorem.
 
 Returning to the non-zero value $\mathbf{j}_{b}(\mathbf{r}_{a}, \mathbf{r}_{b})$ we
may ask why there are  correlations between the detector particle and the atom, when
the latter is far away in the region $I$.  In the conventional theory, Bohr would argue
that there is no sharp separation between the observing instrument and the atoms even
at this late time in the evolution of the process.  If that argument is rejected,
there is no clear way  to answer to this  question.  

ESSW2 are wrong to have attributed
these probability currents to be an artefact of the Bohm model.  They are essential to standard
quantum mechanics because without them we will not get local conservation of probability and thus
are clearly part of the quantum algorithm. This fact cannot be used  to discredit the Bohm
model without, at the same time discrediting standard quantum  mechanics.

Rather the Bohm interpretation actually helps to understand why this current is
non-zero.    As we shall see below, the Bohm interpretation shows
that there is a connection between the device-particle and the atom and that this
connection is provided through the quantum potential.  In this way we give a
mathematical explanation of Bohr's position and shows why this probability current
does not vanish.  Furthermore this potential is essential to understand why the Bohm
trajectories behave as  they do.


\section{The Bohm approach}\label{sec.4}
Let us now go on to discuss the Bohm approach and show in detail how this applies to
 the interferometer shown in figure 1. We will show that there is no disagreement
between the empirical predictions of orthodox quantum mechanics and the Bohm approach
thus supporting the conclusions of Dewdney et al \cite{DHS}.  In this paper we will
clarify their answer in the light of the analysis of the previous section.

Before
going into specific details, it is necessary to make some general remarks, which, we
hope, will clarify the basis for our discussion.  We would like to emphasise that, for
the purposes of this discussion we will follow {\em strictly} the point of view presented in
Bohm and Hiley \cite{BH87} \cite{BH93}.  This approach differs in some significant
details from that used by D\"{u}rr et al \cite{DF} under the title ``Bohmian
Mechanics''.  Bohm and Hiley \cite{BH93} made it very clear that their approach was
not an attempt to return to a mechanistic view of Nature based on classical physics.
Indeed they went further and argued that it was not possible to provide a consistent
{\em mechanical} explanation of quantum processes.  A much more radical view is
necessary as was detailed in chapters 3, 4 and 6 of their book where they showed why
this approach took us beyond such a mechanical picture.  For example, new concepts
such as active and passive information were introduced specifically to account for the
novel features appearing in quantum processes, but these arguments seem to have gone
unnoticed or implicitly rejected.

Naturally the appropriateness of these ideas for physics are open for debate, but to
our knowledge this has not taken place.  Fortunately for the purposes of the article,
the validity of these ideas is unnecessary and we can stick to a simple interpretation
of the formalism. We do not need to use these new notions explicitly in providing a
consistent account of the experiments discussed in the previous section.  
What we will
show, however, is that the use of particle trajectories in quantum mechanics can provide
a consistent account of all possible experiments of the type shown in 
figure (1).

We will start our discussion free from as
many metaphysical assumptions about the underlying quantum process as possible.  Let
us begin by assuming that the present quantum formalism captures the essential
features of a quantum process and that no modification of its mathematical structure
is necessary.  Our task is simply to explore the formalism in a way that is different
from the usual approach and see if this approach will provide any different insights
into the nature of quantum processes in general.   Thus we will not start with any
preconceived notions of what should, or should not, constitute a quantum process. 
Rather we simply assume that there is some objective process and that the wave
function is not merely part of an algorithm or a `tool', but contains further
objective information about the quantum process. 

Our approach begins with the
observation that if we write the wave function in polar form $\Psi = 
R\mbox{exp}[iS]$, and
substitute it into Schr\"{o}dinger's equation, we obtain two conservation
equations\footnote {We will put $\hbar = 1$ for the rest of the paper.}.
The first of these is a conservation of energy equation,
\begin{equation}
    E=\frac{p^{2}}{2m}+V+Q
\end{equation}
This equation follows from the real part of the Schr\"{o}dinger equation, which is easily 
shown to be
\begin{equation}
    \frac{\partial S}{\partial t}+\frac{(\nabla S)^{2}}{2m}+V+Q=0,
\end{equation}
and which, apart from the additional term $Q$, has the same {\it form} as the
Hamilton-Jacobi equation of  classical mechanics.  We will call equation (19) the quantum
Hamilton-Jacobi equation.  Equation (19) then follows from equation (18) if we use the
Hamilton-Jacobi relations
\begin{equation}
    E=-\frac{\partial S}{\partial 
    t}\hspace{1cm}\mbox{and}\hspace{1cm}\mathbf{p}=\nabla S¥
\end{equation}¥

Since equation (18) is a conservation of energy equation, we can interpret $Q$ as 
introducing a new quality of energy, which is absent in classical mechanics.  The
specific form of $Q$, which we call the quantum potential, is given by
\begin{equation}
    Q(\mathbf{r},t)=-\frac{1}{2m}\frac{\nabla^{2}R(\mathbf{r},t)}{R(\mathbf{r},t)}¥¥
\end{equation}
The similarity to the classical equation suggests that we ought to be able to provide
 a classical view of quantum processes.  However as we explained earlier,
an exploration of the properties $Q$ possesses quickly dispels any possibility of a
return to classical mechanics.  We will not be concerned with these properties in this
paper but refer the interested reader to Bohm and Hiley \cite{BH93}. 

It should be emphasised that this potential is not introduced in an 
{\em ad hoc}
manner.  It is already implicit in the Schr\"{o}dinger equation and its presence is
essential to obtain the same statistical results as those obtained from the orthodox
approach. This new quality of energy plays a crucial role in our approach. 

The other
equation, which is derived from the imaginary part of the Schr\"{o}dinger equation, is
exactly the conservation of probability given by equation (5) expressed in the form
\begin{equation}
    \frac{\partial P}{\partial t}+\nabla .(Pv)=0,
\end{equation}
Here we have identified the probability $P$ with $R^{2}$ in the usual way.

As is well known the classical Hamilton-Jacobi equation provides a set of one-parameter
 solutions, which we immediately identify as particle trajectories.  
 When $Q$ is
non-zero in equation (19), we are still able to find a set of one-parameter solutions
of the form
\begin{equation}
    \mathbf{r}(t)=f(R,S,t),
\end{equation}
which we can obtain simply by integrating the subsidiary condition 
$\mathbf{p}=\nabla S$.  This equation 
is also known as the `guidance condition' \footnote{We would like to emphasise that this
can be regarded as a subsidiary condition, which enables us to interpret equation (19)
as the conservation  of energy equation.}.
Equation (19) leads to the central
question ``What is the meaning of these solutions?''  Could these curves be regarded as
some kind of trajectories even though they are in the quantum domain?

The first objection to making such an identification might be thought to arise from the
uncertainty principle.  The
 notion of a trajectory requires the particle to have a simultaneously well-defined
position and momentum, whereas the uncertainty principle states that we cannot measure
position and momentum simultaneously.  Our ability to measure position and momentum
simultaneously does not logically rule out the possibility that the particle 
{\em has} a
well-defined position and momentum.  It could be that there is something intrinsic in
the measuring process that rules out such a possibility.  This is indeed what happens
as is shown in Bohm and 
Hiley \cite{BH93}.
When the particle is coupled to a measuring device, a new quantum potential arises and
it is the appearance of this quantum potential that ensures that the uncertainty principle
is not violated. 

One important, but by no means necessary, argument for retaining the
notion of a trajectory comes from examining situations where $Q$ changes with time. 
Then, for example, as $Q$ approaches zero, the general one-parameter solutions become
identical to the classical particle trajectories in the limit. Thus there can be a
smooth transition from the classical to the quantum domain.  In other words 
as $Q$
increases from zero, the one-parameter curves also change and this change can become
larger as $Q$ becomes larger.  At no point are we forced to abandon the notion of a
trajectory.  This suggests that it may still be possible to retain the notion of a
`particle' even in the quantum domain. We can then explore the consequences of adopting
this proposal simply to see how far it can be meaningfully sustained.

Alternatively
we could give a more general meaning to these curves.  For example, we could imagine a
deeper, more complex process, which is not localised, but extends over a region of
space where the wave function is non-zero.  The curve could then be interpreted as the
centre of this  activity as this process evolves in space.  As $Q$ becomes smaller, the
region over which it is effective becomes smaller so that in the classical limit, a
point-like property is all that we need. This image of the process has certain
attractive features, but at present there is insufficient structure in the mathematics
as it stands to fully justify such a view.

Whatever the situation, one factor is quite
clear.  The conservation of probability implies that if the initial probability
(defined by $R^{2}_{inital}$) corresponds to the initial quantum probability distribution,
then the final distribution taken over all of these curves will be exactly the same as
the final probability distribution calculated from standard quantum mechanics.  Thus
even identifying these one-parameter curves as `particle trajectories' will not
produce any probabilities that are different from those already predicted by the standard
theory.   

In one sense this can be regarded as a weakness of the Bohm approach; it
produces no new results.  On the other hand it should not be forgotten that the
approach re-focussed attention on the EPR correlations and provided the necessary
background from which Bell \cite{JB64} \cite{JB66} was led to his inequality
which gave rise to testable consequences.  Another of its strengths is that many, if not all, of the
puzzling paradoxes of the standard theory disappear as has been clearly shown in Bohm
and Hiley \cite{BH93} and in Holland \cite{PH}.


\section{ Details of `particle trajectoriesÕ}\label{sec.5}
\subsection{Trajectories with detectors $D_{1}$ and $D_{2}$ in place}\label{subsec.5.1}
Let us now turn to consider in detail the one-parameter solutions of equation 
(19), 
which for the present we will regard as providing a set of `quantum particle
trajectories'.  It is straightforward to calculate these curves for an interferometer
of the type shown in figure 1.

 To provide a comprehensive understanding of the
consequences of these trajectories, let us first consider an interferometer in which
the cavity in the path BM$_{2}$ has been omitted. As is clear from the discussion in
section 2, the region of particular interest is where the beams cross 
at $I$.  Here the
wave function is given by equation (8).

\begin{figure}[t]
\includegraphics{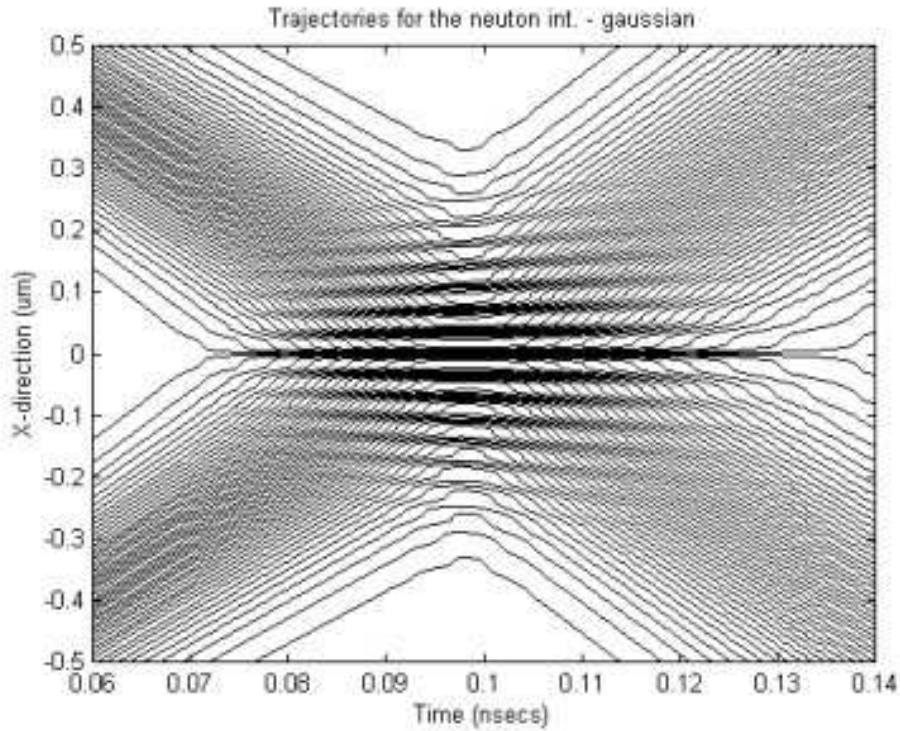}
\caption{Trajectories in the region of $I$.}
\end{figure}

To calculate the trajectories, we must first
write wave function (8) in the form $\Psi = R\mbox{exp}[iS]$ and then find the
expression for $S$, which will be of the form
\begin{equation}
    S(\mathbf{r}_{a},t)=G(R_{1},R_{2},S_{1},S_{2},t).
\end{equation}
We can then use this expression in the subsidiary condition, 
$\mathbf{p}=\nabla S$   to calculate the trajectories.  These are straightforward to 
evaluate numerically. The trajectories in the
region $I$ and its immediate surroundings are shown in figure 2.

\begin{figure}[t]
\includegraphics{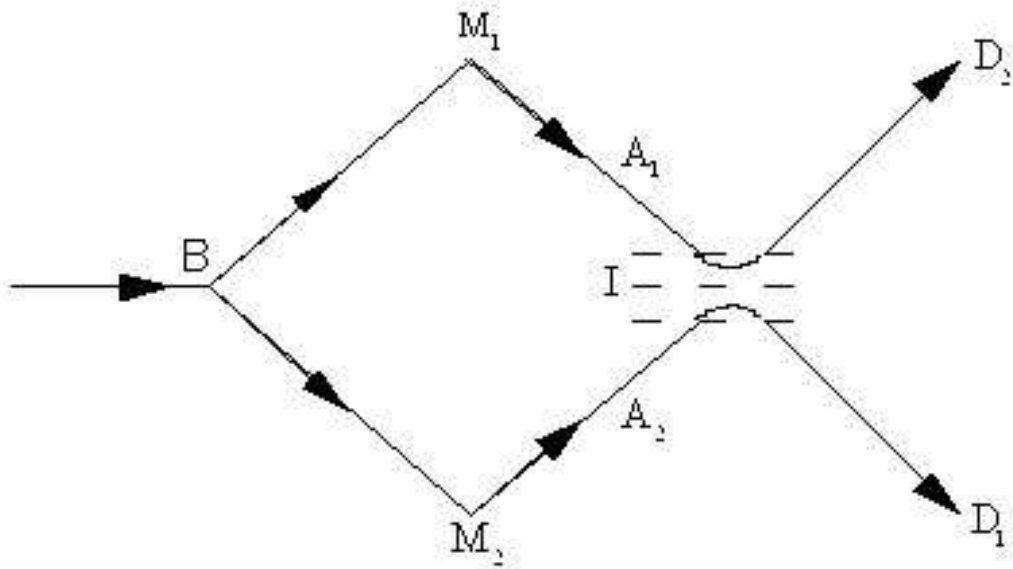}
\caption{Sketch of particle trajectories.}
\end{figure}

The complete trajectories from the beam splitter to the detectors have then been 
sketched  in figure 3 for convenience.  These figures show that the atoms that are
ultimately detected at D$_{1}$ must have travelled along a path 
BM$_{2}$A$_{2}$D$_{1}$, while the atom
that is recorded at D$_{2}$, must travel along a path 
BM$_{1}$A$_{1}$D$_{2}$.  We immediately see here
that the trajectories appear to be reflected about the $z = 0$ plane.  It is this result
that seems to be totally against `common sense' and therefore there must be something
wrong with the Bohm
approach.  However it should be noted that these results are entirely 
 consistent with the quantum
probability currents that we discussed in section 3 where we showed 
that there was no net current crossing the $z=0$ plane.

How is it possible for trajectories to be reflected in the way shown in a
region free of any classical potentials and therefore for no apparent reason?

Actually we do already have a similar type of behaviour in the two-slit interference
experiment \cite{DPH}.   After the particles have passed through the slits,
they no longer follow straight-line trajectories, but show a series of 
`kinks'.  None
of these trajectories cross the horizontal plane of symmetry.  All the particles that
pass through the top slit end up on the top part of the plane of the 
interference pattern.  The
ÔkinksÕ in the trajectories are just sufficient to create the bunching in exactly the
right way to produce the required fringes.  The reason for these kinks was immediately
seen from the calculation of the quantum potential.  This potential
changes rapidly in the region of these kinks and is thus seen to be directly 
responsible for the resulting `interference'.

\begin{figure}[t]
\includegraphics{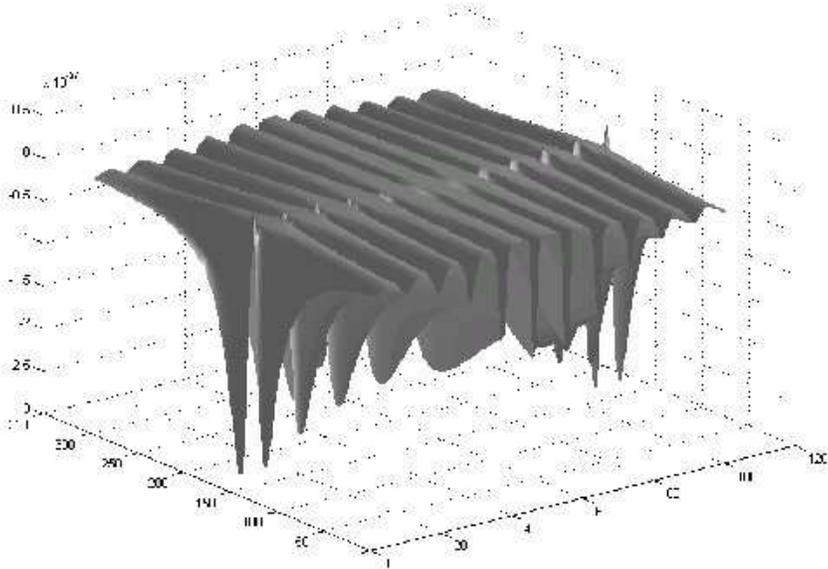}
\caption{Quantum potential in region $I$.}
\end{figure}

We can show that a similar quantum potential is responsible for the behaviour of the 
trajectories crossing region $I$ in the interferometer we are considering here.  We can
calculate this quantum potential $Q$, using the wave function $\psi 
(\mathbf{r}_{a})$ to obtain an
expression for the amplitude $R$
\begin{equation}
    R(\mathbf{r}_{a},t)=F(R_{1},R_{2},S_{1},S_{2},t)¥¥¥¥¥
\end{equation}
The result of the calculation for $Q$ is shown in figure 4.

 This behaviour is exactly what we would expect in a region where the wave functions
overlap. A close examination of the details of the potential shows that it exactly
accounts for the shape of the trajectories shown in figure 2.  In this way we have an
explanation of why the Bohm trajectories are reflected in the $z = 0$ plane and we have
a causal explanation of why the trajectories behave as they do.  It is this feature
that  leads to the conclusion that `Bohm trajectories do not cross'.


\subsection{Trajectories with the cavity in place}\label{subsec.5.2}
        As we have seen the serious challenge made by ESSW2 arises when the micromaser cavity
 is added to one of the arms of the interferometer as shown in figure 1.   To discuss
the consequences of adding this cavity for the Bohm trajectories, we can simplify the
problem considerably by following Dewdney et al \cite{DHS} and replacing the cavity by a
particle in a one-dimensional `box' described by two wave functions, 
$\Phi _{0}(\mathbf{r}_{b})$ for the
unexcited state and $\Phi _{E}(\mathbf{r}_{b})$ for the excited 
state\footnote {The Bohm approach can be applied to the field in the cavity as has been 
discussed by Bohm, Hiley and Kaloyerou \cite{BHK}, Bohm and Hiley 
\cite{BH93} and Kaloyerou \cite{K}.  
The details of the application to the cavity in relation to this situation will be published 
elsewhere (See also Lam and Dewdney \cite{LD}.}.
 Here $\mathbf{r}_{b}$ is the  position co-ordinate of
the particle in the box.  We will continue to assume that the wave functions of the
atom, $\psi_{1}(\mathbf{r}_{a}, t)$ and $\psi_{2}(\mathbf{r}_{a}, t)$, are Gaussians of small width.

The wave function for
the system at time $t_{1}$ after the atom has sufficient time to interact with 
the `cavity'\footnote{We will continue to call the two-state system a `cavity'.},
but not yet had time for their the Gaussian wave packets to  overlap (i.e., they 
have not yet reached the  region $I$)
will be either
\begin{equation}
    \Psi (\mathbf{r}_{a},\mathbf{r}_{b},t_{1})=\psi_{1}
    (\mathbf{r}_{a},t_{1})\Phi_{0}(\mathbf{r}_{b})
\end{equation}
or
\begin{equation}
     \Psi (\mathbf{r}_{a},\mathbf{r}_{b},t_{1})=\psi_{2}
    (\mathbf{r}_{a},t_{1})\Phi_{E}(\mathbf{r}_{b})
\end{equation}
By using the method described in section 3.1, the set of trajectories centred on
BM$_{1}$A$_{1}$ can be calculated using equation (26), while those centred 
on BM$_{2}$A$_{2}$ can be
calculated from equation (27). 

We can check that these give the expected outcome by
moving the detectors D$_{1}$ and D$_{2}$ to the positions A$_{1}$ and 
A$_{2}$ (See figure 5 below).  We
will then be able to confirm that the atom that goes through the `cavity' will be
recorded D$_{2}$, while the atom that does not go through the `cavity' will be recorded at
D$_{1}$.  In the first case, the energy of this atom should, of course, be less since it
has exchanged energy with the `cavity', and this can be checked by putting an
energy-measuring device, D$_{4}$, at A$_{2}$.   All of this is exactly as we would expect and no
strange or unacceptable behaviour results at this stage. 

As we have seen the problem
arises once we allow the Gaussian wave packets to overlap again in the region of
interference $I$.  Since we have assumed that there is no coherence loss when the atom
has passed through the `cavity', the wave function must now be written in the form
\begin{equation}
    \Psi (\mathbf{r}_{a},\mathbf{r}_{b},t_{1})=\psi_{1}
    (\mathbf{r}_{a},t_{1})\Phi_{0}(\mathbf{r}_{b})+\psi_{2}
    (\mathbf{r}_{a},t_{1})\Phi_{E}(\mathbf{r}_{b})
\end{equation}
Let us now examine what happens in the region of interference $I$ in this case.  The
 wave functions $\Phi _{0}(\mathbf{r}_{b})$  and $\Phi _{E}(\mathbf{r}_{b})$  
 are both real so that equation (28) can be written
in the form
\begin{equation}
    \Psi_{f}(\mathbf{r}_{a},\mathbf{r}_{b}, t)=R_{1}(\mathbf{r}_{a},\mathbf{r}_{b}, t)
    exp[iS_{1}(\mathbf{r}_{a}, t)]+R_{2}(\mathbf{r}_{a},\mathbf{r}_{b}, t)
    exp[iS_{2}(\mathbf{r}_{a}, t)]
\end{equation}
In order to calculate the trajectories, we must again write this wave function in 
the form
\begin{equation}
    \Psi_{f}(\mathbf{r}_{a},\mathbf{r}_{b}, t)=R(\mathbf{r}_{a},\mathbf{r}_{b}, t)
    exp[iS(\mathbf{r}_{a},\mathbf{r}_{b}, t)]
\end{equation}
We immediately see an important difference between this case and the case described 
by equation (23).  Here $R$ and $S$ are functions of both $\mathbf{r}_{a}$ 
and $\mathbf{r}_{b}$, rather than $\mathbf{r}_{a}$ alone
and therefore we have {\em a pair of coupled one-parameter solutions} of the real part of
the Schr\"{o}dinger equation.  These are given by
\begin{equation}
    \mathbf{p}_{a}=\nabla_{{\mathbf{r}_{a}}}S(\mathbf{r}_{a},\mathbf{r}_{b}, t)
    \hspace{1cm}\mbox{and}\hspace{1cm}
    \mathbf{p}_{b}=\nabla_{{\mathbf{r}_{b}}}S(\mathbf{r}_{a},\mathbf{r}_{b, t})
\end{equation}
This means that as the atom moves along its trajectory in the region of interference,
 $I$, the particle in the box also moves, showing that this particle is still coupled to
the atom even though they are separated in space.  This would then also account for why the
probability current for the `particle in the box' (the cavity) discussed in section 3 is different from
zero as the atom passes through the region $I$.

From the classical point of view this
behaviour would be absurd. However when we examine this more closely, we find that it is the
quantum potential that mediates this coupling, and recall  {\em this coupling is a necessary
consequence of the Schr\"{o}dinger equation and is not an arbitrary feature imposed from
the outside to satisfy some metaphysical pre-requisite}\footnote{Recall the Bohm approach is
driven by the quantum formalism and not by any pre-assumed metaphysics.}.

\begin{figure}[t]
\includegraphics{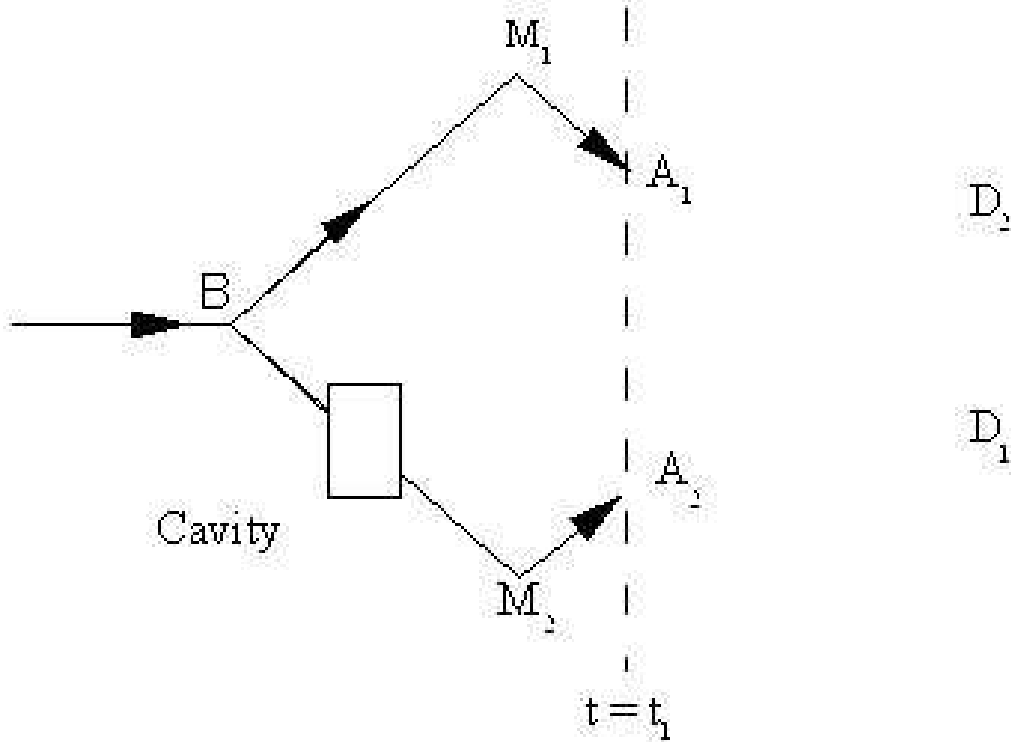}
\caption{Trajectories at time $t = t_{1}$.}
\end{figure}

What makes this quantum potential seem particularly unpalatable is its gross 
non-locality or non-separability.  Here we have the surprising feature that the atom
and the `cavity' are still coupled long after the atom should have passed (or not
passed) through the `cavity'. The process is not complete once the wave packet has
passed through the cavity. This is a very clear example of why  Wheeler argues that the
process is ``brought to a close by an  irreversible act of amplification''.

However this behaviour should not be so surprising, it is
exactly the situation found in the EPR paradox \cite{BH75} and, indeed in quantum 
teleportation \cite{MH}. 
It is the quantum potential that provides an explanation of these effects and one can
see that this coupling is essential to conserve energy. (For further details of this
point see Hiley \cite{BJH99}).

Let us now go on to examine the trajectories of the atom in
detail. Four sets of trajectories for particular sets of initial 
conditions $\{\mathbf{r}_{a_{0}}\}$,
and for four different values of $\mathbf{r}_{b_{0}}$ are shown in
figure 6.  These trajectories exhibit `wobbles' that are a typical signature for
interference-type behaviour and these are a necessary consequence of the coherence
between the two wave packets. 

Does this mean that the Bohm approach predicts
interference in the region $I$?  If this were the case, the Bohm interpretation would
clearly disagree with the standard interpretation, which predicts that there are no
visible interference effects in the region $I$ and indeed no fringes 
are visible in this region.

As we have already shown in section 2,
the wave function is given by equation (1) which gives no interference fringes because
 $\Phi _{0}(\mathbf{r}_{b})$ and  $\Phi _{E}(\mathbf{r}_{b})$ are orthogonal. 
 In other words integrating over all  $\mathbf{r}_{b}$, destroys the interference
terms.  This integration over $\mathbf{r}_{b}$ provides the clue as to why the Bohm 
approach does
not predict any interference effects either.  Although each particular set of trajectories
shown in figure 6 do show interference-type `wobbles', they do so only for the curves
calculated for a given $\mathbf{r}_{b_{0}}$. 

 When we average over the ensemble of
trajectories over different initial $\mathbf{r}_{b_{0}}$, no interference effects
appear  in the region $I$.  This is
because the set of positions of this ensemble at some time $t = t_{2}$ when the atoms
would
 be
in the region, $I$, show a uniform distribution, rather than a fringe pattern. 
Thus the
total of all Bohm trajectories do not bunch to form an interference pattern so there
is no disagreement with quantum mechanics on this point.

\begin{figure}
\includegraphics{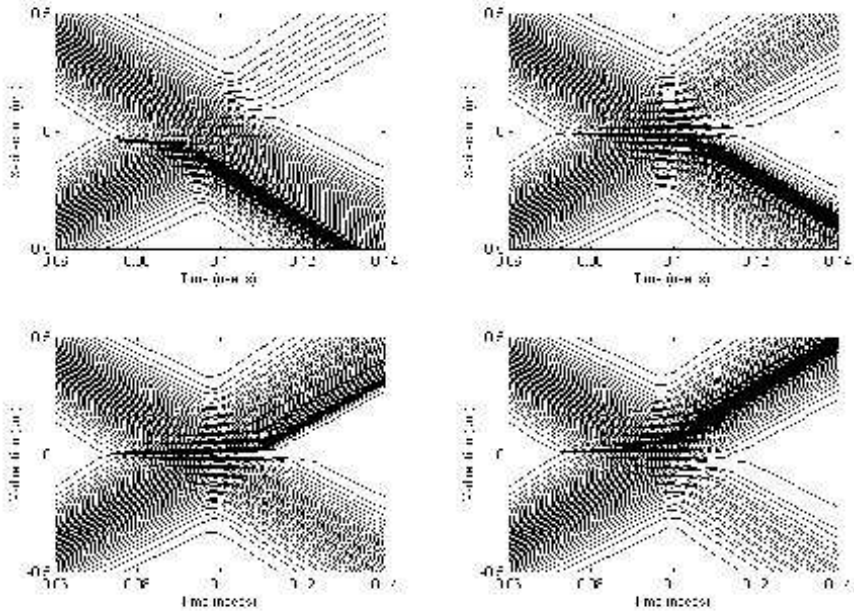}
\caption{Trajectories for four different $\mathbf{r}_{b0}$.}
\end{figure}

Now we come to the crucial
point from which the objections have been raised.  The calculations for 
different $\mathbf{r}_{b}$
show that a significant number, although by no means all, of the trajectories will be
similar to those illustrated in figure 4.  That is they will be `reflected' in the
region of overlap and it is the presence of this type of trajectory that led ESSW2 to the
conclusion that the trajectories must be rejected because of their bizarre behaviour. 
Their argument runs as follows. 

Suppose an atom follows the trajectory BCM$_{2}$A$_{2}$D$_{1}$.  On
passing through C it gives up energy to the `cavity', as we have already seen when we
discussed what happened when we measured the energy of an atom at A$_{2}$.  But according
to the trajectory picture any atom on that path would have ended up in the detector
D$_{1}$.  

If we were to measure the energy of this atom just prior to entering D$_{1}$, we would
find that it had NOT lost any energy.   Indeed its wave function is 
$\psi_{1}(\mathbf{r}_{a})$ which
indicates it must have the same energy as when it entered the beam-splitter B.   There is
no loss of energy because if we also measure the energy of the `cavity' at any time
after the atom had left region $I$, we would have found that it had NOT gained any
energy!    Yet any energy measurement prior to the atom reaching the region $I$ would
show that the atom had lost energy to the `cavity' and the `cavity' had gained
energy.  It is this feature that is very surprising and which is used to suggest that the Bohm
approach is flawed. 

Of course a similar argument can also be applied to an atom that
ends up at D$_{2}$ after following the trajectory BM$_{1}$A$_{1}$.  In this case a
measurement of its energy just before it arrived at D$_{2}$ would show that it had lost
energy to the `cavity' even though it had not been anywhere near the `cavity'.  How
could this possibly happen?

Clearly if an atom that goes through a `cavity' without
exciting it, or conversely, if an atom that does not go through the `cavity' succeeds
in exciting it without any apparent connection between atom and 
`cavity', must be
regarded as behaving `unreasonably'.  In this case we would be forced to conclude that
the trajectories do not have any physical meaning. However the crucial phrase is
``without any apparent connection between atom and cavity''.  

As we have already
explained there is a `connection' between the atom and the `cavity'.  The connection
appears in the real part of the Schr\"{o}dinger equation itself where it takes the form of
the quantum potential.  ESSW2 have completely ignored this aspect of the Bohm
ontology.  For as Bohm and Hiley \cite{BH93A} point out, one of the key features of the
ontology is that ``This particle is never separate from a new type of field that
fundamentally affects it.  This field is given by $R$ and $S$ or alternatively 
by $\psi = R\mbox{exp}[iS]$. $\psi$ then satisfies Schr\"{o}dinger's equation (rather
than, for example, Maxwell's equation), so that it too changes continuously and is
causally determined.''  

Thus the quantum potential is an essential part of the description. 
Without taking the causally determined field into account and only giving relevance to
the trajectories derived from the guidance condition, it is not surprising that the
resulting behaviour has been regarded as `unacceptable'.


\subsection{ Detailed account of the Bohm trajectories}\label{subsec.5.3}
Consider the atom as it moves along the path BCM$_{2}$A$_{2}$.  As it passes through the
 `cavity', an interaction Hamiltonian couples the wave function representing the atom
to wave function of the `cavity' causing it to change, as well as inducing a
corresponding change in the wave packet of the atom.  After the interaction has
finished, the wave function of the `cavity' (in this case the particle in the box) is
real.  The Bohmian approach then shows that the particle in the box is stationary and
in consequence any excitation energy of the `cavity' is stored entirely as quantum
potential energy.   

As the atom passes through the interference region $I$, a new
quantum potential energy is generated.  It must be emphasised that since this energy
arises from equation (25) using equation (1), it must of necessity include the quantum
potential energy stored in the `cavity'.  It is this coupling that gives rise to any
exchange of energy between the `cavity' and the atom so that when the atom emerges
from the region
$I$, it has regained its original energy and the `cavity' is no longer excited.  In
other words the process has been truly `erased'. 

If the particle follows the other
route, the `cavity' is not excited until the particle reaches the region of
interference.  Here the wave packet carrying information about the `cavity' comes into
effect and energy, in the form of quantum potential energy, is again redistributed so
that the cavity becomes excited and the atom loses energy if it is travelling along
one of the `reflected' trajectories.  

Notice that no {\em external} energy is involved in
this process.  It is merely a re-distribution of {\em internal} energy of the two systems linked by
the wave function (1).  This is merely another way of demonstrating  what Bohr 
\cite{NB61B}  called the
`wholeness of the phenomenon'. The two spatially separated systems still form a totality until an
irreversible change takes place.  After this change the two systems 
become independent uncoupled systems.

If we look at this behaviour from the standpoint of
classical physics of course the explanation seems bizarre.  The classical particle is
the centre of the activity and all energy is either kinetic energy or the potential energy
arising from an interaction with some external system.  In this case all energy
exchanges must occur only through a local interaction between the particle and any
externally applied force.   

Quantum phenomena have an inner structure that cannot be
sharply divided into separate sub-systems interacting only through classical forces
described mathematically by a Hamiltonian.  However if we do separate a system into
sub-systems, as we do in the Bohm approach, then it is necessary to have some feature
that reflects this relationship of `indivisibility' between these sub-systems and this
feature is provided by the quantum potential \cite{BH81}.  Thus the quantum
potential reflects the essential `non-separability' or `wholeness' of quantum
processes that Bohr recognised to lie at the heart of quantum processes.  This is the
reason why the quantum potential plays an essential role in the Bohm approach.

In the
above analysis we are forced to attribute new properties to the particle and to the
quantum potential that are totally different from those associated with classical
particles and classical potentials.  If one refuses to recognise the need for these
novel properties, which we regard to be essential to obtain a consistent
interpretation of the formalism, then it is very easy to ridicule the approach.  All
one has to do is to contrast these properties with those of classical physics and to
conclude there are `unacceptable' and `surreal'\footnote {It is interesting to note 
that the surrealist movement in art claimed that there was more to reality than mere 
outward manifestations.  There was a deeper reality (literally ÔsurrealÕ means Ôsuper
realityÕ) that lay behind outward appearances.  When the word ÔsurrealÕ is used with 
its intended meaning, then Ôsurreal trajectoriesÕ is the correct term to describe them! 
Unfortunately ESSW2 use the term in a pejorative sense.}.


\subsection{Replacement of Cavity by an Energy Measuring Device 
D$_{3}$}\label{subsec.5.4}
In order to complete the technical side of the discussion, let us replace the 
`cavity' with a ``position detector'' D$_{3}$ as shown in figure 7 below.  We will assume
this detector has 100$\%$ efficiency and let us further assume, for the present, that the
atom emerging from D$_{3}$ can be described by a wave function 
$\psi_{2}(\mathbf{r}_{a})$¥, which is still
coherent with $\psi_{1}(\mathbf{r}_{a})$.  The wave function in this case would be
\begin{eqnarray}
    \Psi(\mathbf{r}_{a},\mathbf{r}_{c},\mathbf{r}_{d},t_{3})=
    \psi_{1}(\mathbf{r}_{a},t_{3})\Lambda_{unfired}(\mathbf{r}_{c},t_{3})\Omega_{unfired}
    (\mathbf{r}_{d},t_{3})\nonumber\\
    +\psi_{2}(\mathbf{r}_{a},t_{3})\Lambda_{fired}(\mathbf{r}_{c},t_{3})\Omega_{fired}
    (\mathbf{r}_{d},t_{3})
    \label{eq:29}
\end{eqnarray}
where $\Lambda_{i}(\mathbf{r}_{c}, t_{3})$ are the wave functions of the 
detector D$_{3}$ and $\Omega_{i}(\mathbf{r}_{d},t_{3})$ are the
 wave functions of detector D$_{2}$\footnote {For simplicity we assume the detector can
 be described by a wave function.  Expressing it in terms of a density matrix does not
change 
 any principles involved.}.

\begin{figure}
\includegraphics{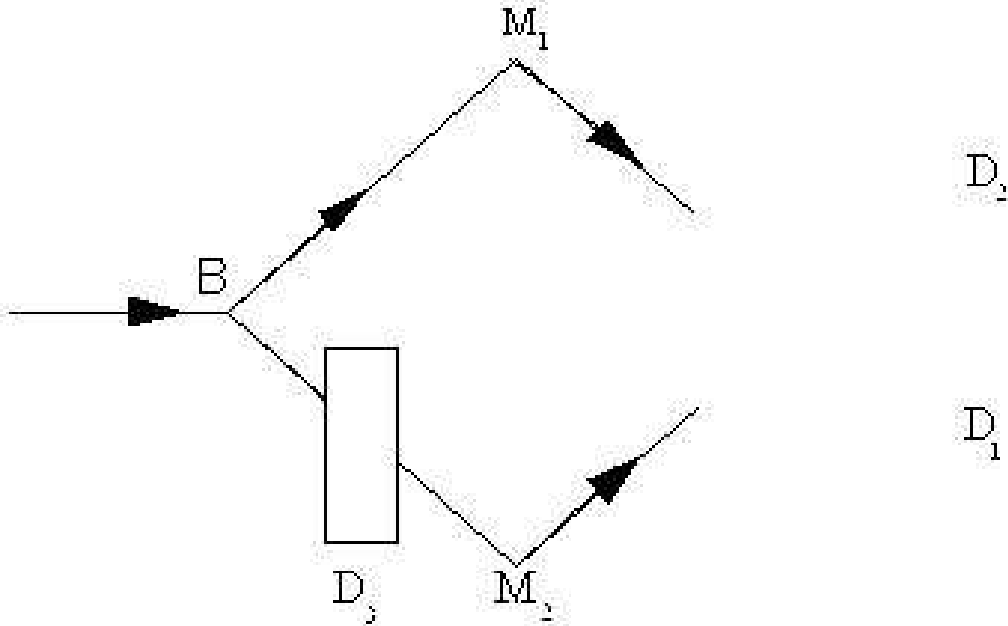}
\caption{Interferometer with position detector D$_{3}$ in place.}
\end{figure}

Here one could argue that because of the rule ``Bohm trajectories do not cross'', we
 will still obtain the odd behaviour of the `trajectories' of the type shown in figure
2 as they cross the region $I$.  For example, we know from the wave function 
(1) that
D$_{2}$ will always fire after D$_{3}$ has fired.  The `no-crossing' rule for Bohmian
trajectories would suggest that the atom should travel along the path 
BM$_{1}$D$_{2}$.  This
would mean that D$_{3}$ had fired even though the particle had not gone 
through D$_{3}$. This
looks as if we will have a situation similar to the case of the `cavity'. 

However
this is not correct.  On detection in D$_{3}$ an irreversible record is left in the
device.   If we were forced to use the explanation that in this case we still need an
exchange of nonlocal energy, then we would see this `irreversible' mark disappear from
the recording of the detector D$_{3}$ as the atom passes through the 
region $I$.  This would,
indeed, be totally unacceptable and would stretch the plausibility of the assumption
that the one-parameter solutions of equation (19) are particle trajectories. 
Fortunately this situation does {\em not} occur in the Bohm interpretation.

To show this let
us consider the wave function after the particle has sufficient 
time, $t = t_{2}$, to pass
M$_{1}$ or M$_{2}$, but before it reaches the detector D$_{2}$
\begin{eqnarray}
    \Psi(\mathbf{r}_{a},\mathbf{r}_{c},\mathbf{r}_{d},t_{3})   
= [\psi_{1}(\mathbf{r}_{a},t_{3})\Lambda_{unfired}(\mathbf{r}_{c},t_{3})\nonumber\\
    +\psi_{2}(\mathbf{r}_{a},t_{3})\Lambda_{fired}(\mathbf{r}_{c},t_{3})]\Omega_{unfired}
    (\mathbf{r}_{d},t_{3})
\end{eqnarray}                          
We can again ask what happens in region $I$.  Let us begin by calculating the quantum
 potential in the region $I$.  Will it still contain interference terms or not? 

To answer this question we must first write
\begin{equation}
    \Psi_{final}=R_{f}(\mathbf{r}_{a},\mathbf{r}_{c},t_{2})
    exp[iS_{f}(\mathbf{r}_{a},\mathbf{r}_{c},t_{2})]
\end{equation}
with
\begin{equation}
    R_{f}(\mathbf{r}_{a},\mathbf{r}_{c},t_{2})=K[R_{uf1}(\mathbf{r}_{a},\mathbf{r}_{c},t_{2}),
    R_{f2}(\mathbf{r}_{a},\mathbf{r}_{c},t_{2}),S_{uf1}(\mathbf{r}_{a},\mathbf{r}_{c},t_{2})
    S_{f2}(\mathbf{r}_{a},\mathbf{r}_{c},t_{2})]
\end{equation}
and
\begin{equation}
    S_{f}(\mathbf{r}_{a},\mathbf{r}_{c},t_{2})=L[R_{uf1}(\mathbf{r}_{a},\mathbf{r}_{c},t_{2}),
    R_{f2}(\mathbf{r}_{a},\mathbf{r}_{c},t_{2}),S_{uf1}(\mathbf{r}_{a},\mathbf{r}_{c},t_{2})
    S_{f2}(\mathbf{r}_{a},\mathbf{r}_{c},t_{2})]
\end{equation}
From equation (35), we can evaluate the quantum potential acting on the particle using
\begin{equation}
    Q_{f}(\mathbf{r}_{a},\mathbf{r}_{c},t)=-\frac{1}{2m}
    \frac{\nabla_{\mathbf{r}_{a}}^{2}R_{f}(\mathbf{r}_{a},\mathbf{r}_{c},t)}
    {R_{f}(\mathbf{r}_{a},\mathbf{r}_{c},t)}
\end{equation}
This must be evaluated at the final positions of the set of values of 
$\{\mathbf{r}_{c}\}$.  

If D$_{3}$ is
 to be a measuring device, then the position of the two sets of variables for the
fired and unfired states must be sufficiently different so that the two final states
of D$_{3}$ can be clearly distinguished.  This is equivalent to requiring the wave packets
describing the two possible D$_{3}$ states not to be overlapping in the variables
$\{\mathbf{r}_{c}\}$.
Because of this requirement, if D$_{3}$ does not fire, the contribution to the quantum
potential $Q_{f}$¥ will only come from $R_{uf1}$ and $S_{uf1}$.  
The other two terms, $R_{f2}$ and $S_{f2}$ will
not contribute to $Q_{f}$ because they are zero when the set 
$\{\mathbf{r}_{c, unfired}\}$ is substituted in
to equation (32).  

On the other hand if D$_{3}$ does fire, then the contributions 
to $Q_{f}$
will only come from $R_{f2}$ and $S_{f2}$, because the other two terms will be zero when
evaluated at the positions set $\{\mathbf{r}_{c, fired}\}$.

\begin{figure}[t]
\includegraphics{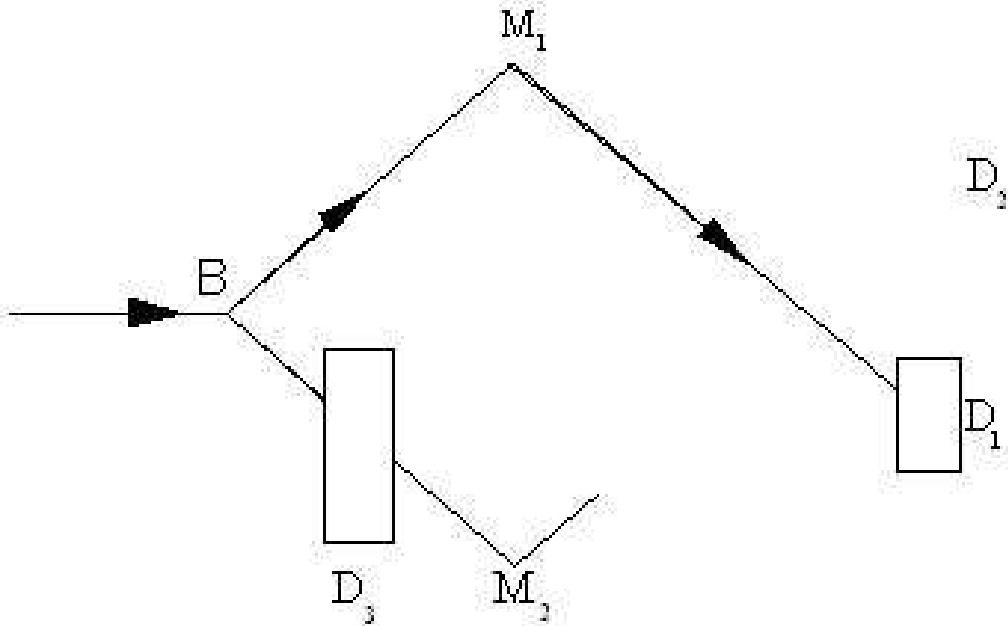}
\caption{Detector D$_{3}$ does not fire.}
\end{figure}

Thus $Q_{f}$ will never contain contributions from the path that the particle did not take.
  In consequence there will be no interference terms so the two possible paths will be

1.  BM$_{1}$D$_{1}$ if D$_{3}$ does not fire so that D$_{1}$ does fire. (See figure 8)

2.  BD$_{3}$M$_{2}$D$_{2}$ if D$_{3}$ does fire.  In this case D$_{2}$ will fire. (See figure
9)

\begin{figure}
\includegraphics{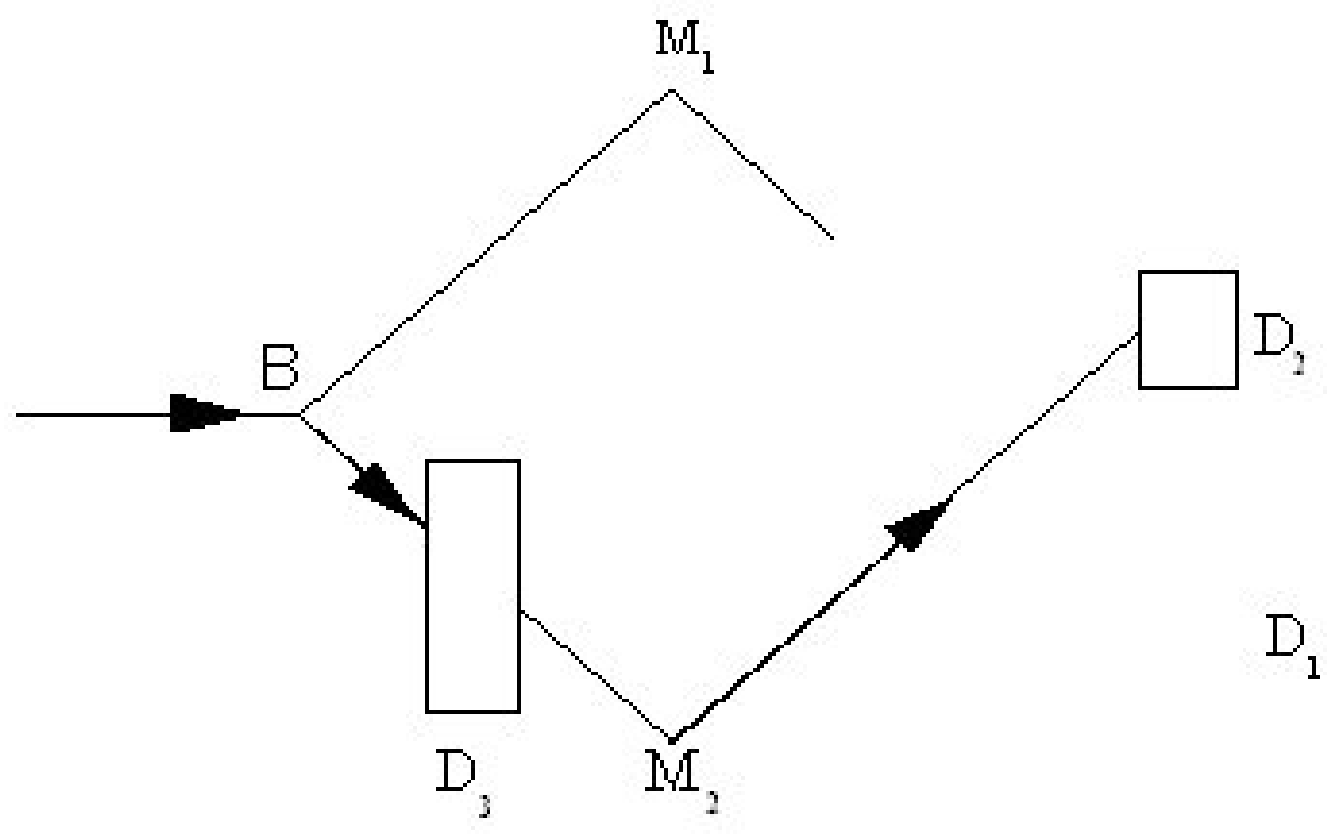}
\caption{Detector D$_{3}$  fires.}
\end{figure}

The particle trajectories are therefore straight lines from M$_{1}$ to 
D$_{1}$ if D$_{3}$ does not 
fire, or straight lines from M$_{2}$ to D$_{2}$ if D$_{3}$ does fire. Thus both sets of trajectories
pass straight through the region $I$ without showing any coherence.  This is exactly the
situation described by the density operator
\begin{equation}
    \rho 
    =|\psi_{1}\rangle|\Lambda_{unfired}\rangle \langle\psi_{1}|\langle 
    \Lambda_{unfired}|+
    |\psi_{2}\rangle|\Lambda_{fired}\rangle \langle\psi_{2}|\langle 
    \Lambda_{fired}|
    \label{eq:35}
\end{equation}
The consequences of this density operator are the same as those described in section 2.

In this case the rule that ``Bohm trajectories do not cross'' is not violated because 
the relevant space in which the non-crossing rule works is {\em configuration space}. The
introduction of the detector D$_{3}$ increases the number of dimensions of the
configuration space to include the set $\{\mathbf{r}_{c}\}$ of relevant detector particles.  What this
means is that each trajectory is parameterised by $\mathbf{r}_{a}$ and  $\{\mathbf{r}_{c}\}$. 
Since the set $\{\mathbf{r}_{c,fired}\}$
 and the set $\{\mathbf{r}_{c,unfired}\}$ are distinct, the set of trajectories
 corresponding to
when  D$_{3}$ fires is distinct from the set when D$_{3}$ does not fire and 
therefore the
trajectories corresponding to each situation do not cross in the higher dimensional
configuration space.  The trajectories only appear to cross when they are projected
into the two-dimensional ($\mathbf{r}_{a}$ ) configuration space.  Thus there is no violation of the
``no crossing'' rule. 

This completes the detailed description of the Bohm trajectories
in the various possible structures that arise in the type of interferometer
arrangements discussed by ESSW2, ESSW3 and Scully \cite{S}.
\subsection{The criticisms of Aharonov and Vaidman}\label{subsec.5.5}
Finally we want to consider the criticisms of Aharonov and Vaidman 
\cite{AV}.  Although
 they are mainly to do with a model that is fundamentally different from that
considered in this paper, they have drawn attention to a problem that 
is
perceived to present a problem for the Bohm approach we are using here. 

The problem
involves replacing the cavity with a bubble chamber and replacing the atoms with
particles that can ionise the liquid molecules in the bubble chamber. Aharonov and
Vaidman argue that 
\begin{quote}
    The bubbles created due to the passage of the particle are
developed slowly enough such that during the time of motion of the particle the
density of the spatial wave function of each bubble does not change significantly.
\end{quote}
In this case they argue that we will see the trace of the bubbles behaving as if the
particle moves in one arm, whereas it actually travels down the other arm.

We will now
show that this does not happen.  To do this we must be much more careful in analysing
the process that is responsible for the bubbles forming in the first place.  The
starting point for the development of a bubble is an ionised atom or molecule that
provides the nucleus for a bubble to form in the first place.  Let us simplify the
discussion by discussing the interaction of the particle with one atom of the bubble
chamber liquid. Let the wave function of the un-ionised atom be 
$\Psi_{UA}(\mathbf{r}_{A},\mathbf{r}_{e})$ where $\mathbf{r}_{A}$ is
the centre of mass co-ordinate of the atom and $\mathbf{r}_{e}$ is the position of 
the electron that
will be ejected from the atom.  After the ionising interaction has taken place, the
wave function of the ionised atom will be $\Psi_{IA}(\mathbf{r}_{A})$ and the wave function of the ejected
electron will be $\phi (\mathbf{r}_{e})$.  The total wave function at time 
$t =t_{1}$ will be
\begin{equation}
    \Psi(\mathbf{r}_{a},\mathbf{r}_{A},\mathbf{r}_{e},t_{1})=
    \psi_{1}(\mathbf{r}_{a})\Psi_{UA}(\mathbf{r}_{A},\mathbf{r}_{e})+
    \psi_{2}(\mathbf{r}_{a})\Psi_{IA}(\mathbf{r}_{A})\phi (\mathbf{r}_{e})
\end{equation}
Before the bubble can begin to form on the ionised atom, the electron must be 
removed a sufficient distance so that the probability of finding the electron at the
atom is zero.  This means that the wave function of the ionised atom does not overlap
with the wave function of the electron.  We can now write this wave function in the
form
\begin{equation}
     \Psi_{final}=R_{f}(\mathbf{r}_{a},\mathbf{r}_{A},\mathbf{r}_{e},t_{1})
    \mbox{exp}[iS_{f}(\mathbf{r}_{a},\mathbf{r}_{A},\mathbf{r}_{e},t_{1})]
    \label{eq:37}
\end{equation}

We can then evaluate the quantum potential and find that it contains no interference 
terms.  Once again the reason for this is that the expression must be evaluated for
the positions of {\em all} the particles. 

If the particle does not ionise the molecule,
ie, it travels on the path BM$_{1}$A$_{1}$ then the contribution of the second term will
be zero
because the electron will be in the position where $\phi (\mathbf{r}_{e})$ is zero. 
On the other hand
if the particle does ionise the atom, the first term will give no contribution to
the quantum potential because $\Psi_{UA}(\mathbf{r}_{A},\mathbf{r}_{e})$ is zero for 
the ionised position of the
electron.  This means that when ionisation is involved, the wave function 
(39)
effectively behaves like the density operator as far as the Bohm approach is
concerned.  In other words the two paths do not produce any interference effects, so
the particle either follows the path BM$_{1}$A$_{1}$D$_{1}$ or 
BM$_{2}$A$_{2}$D$_{2}$.  Thus the particle that ionises the atoms of the liquid actually pass
through the points that eventually develop into the bubble track.  It does not matter
how slowly the bubbles develop, it is the ionisation process that destroys any
interference effects in the cross-over region $I$.  Thus the Bohm approach when
evaluated correctly does not give the results claimed for it by Aharonov and Vaidman
\cite{AV}.
 

\section{Conclusions}\label{sec.6}
In this paper we have shown that the claim that we can meaningfully talk about
 particle trajectories in an interferometer such as the one shown in figure 1 within
quantum mechanics made by ESSW2 \cite{ESSW2} and by Scully \cite{S} does not follow 
from the standard (Copenhagen) interpretation. An additional assumption must be made,
namely, that the cavity and the atom can only exchange energy when the 
atom {\em actually}
passes through the cavity.  Here the position of the particle becomes an additional
parameter, which supplements the wave function and therefore is not part of the
orthodox interpretation.  Furthermore we have shown that this way of 
introducing the position coordinate leads to a contradiction
as we are forced to conclude that although the atom follows one path, it behaves as if
it went down both paths.

In standard quantum mechanics, inferences about the `path'
of a particle can only be made in terms of a series of measurements.  As Heisenberg
\cite{WH} states  ``By path we understand a series of points in space which the electron
takes as `positions' one after another.''  He adds:  ``When one wants to be clear about
what is to be understood by the words `position of the object', then one must specify
definite experiments with whose help one plans to measure the `position of the
electron'; otherwise this word has no meaning''.

        The claim by Scully \cite{S} is that the cavity constitutes a potential
 position-measuring device cannot be sustained.  A measuring device giving rise to an
observation requires, according to Bohr \cite{NB61}, some form of amplification that
involves an irreversible process. An exchange of energy with the cavity 
{\em per se} does
not involve any amplification or irreversible process and therefore the cavity does
not constitute a position-measuring device in the sense assumed in the orthodox
interpretation of quantum mechanics.  This means that the addition of the cavity to
one arm of an interferometer cannot be used as a {\em Welcher Weg}  (WW)device from within
the Copenhagen interpretation.

Of course, the subsequent measurement of the energy of
the cavity does involve amplification and irreversibility.  But as we have seen in
section 2, the cavity cannot be taken to be a position-measuring device once the atom
has passed through the region of interference {\em I} without an additional assumption,
which we have shown leads to the contradiction discussed above. 

Even when discussing
energy exchanges, the orthodox interpretation insists on using only the wave function,
without reference to any additional parameters such as the particleÕs actual position. That is the
interaction Hamiltonian only introduces changes in the {\em wave functions} of the particle
and the cavity {\em independently of the actual position of the particle}.
 Recall the quotation of Bohr \cite{NB61A} given in section 2.1 where he
points out that the impossibility of sub-dividing quantum phenomena leads to an ambiguity in
attributing physical properties to atomic objects and this ambiguity demands a description being given
{\em only in terms of wave functions}. Thus not only is it unnecessary to insist that the particle must
have been through the  cavity, {\em it actually has no meaning to make such a claim}. 

Indeed as Bohr points out, we cannot sub-divide a quantum process and analyse it in the way we do
in classical physics.  All we can rely on is the well-defined quantum algorithm from which we are able
to calculate the probabilities of a given outcome defined by the whole experimental
arrangement \cite{NB61}.   This process does not attach any notion of reality to the
wave function; it is regarded simply as part of the mathematical algorithm, or if you
will a `tool', giving it no ontological status at all.  It is merely a symbol in a
mathematical formula and as long as we adopt this position we cannot talk about which
way particles go when interference is involved. If we want to give direct significance
to the particle and remain in total agreement with the quantum formalism then we must
use the Bohm approach.

In fact in section 5 we have provided a complete account of the Bohm
interpretation as applied to the interferometer introduced by ESSW2 and Scully.
We have shown that the observational consequences of the Bohm approach are exactly the
same as those obtained from the orthodox interpretation when the cavity 
is in place.  To prevent any further misunderstanding of the 
consequences of the Bohm approach we will summarise the results in the 
various situations involving the interferometer shown in figure 1.
 
If energy measurements are made at a time $t = t_{1}$ (see figure 6) by inserting the
device at A$_{1}$ or A$_{2}$ then

1. The atom arriving at A$_{1}$ will be excited and no energy will be detected in the
 cavity.
 
2. The atom arriving at A$_{2}$ will have lost its energy and the cavity will be excited.

If the energy measurements are made at time $t = t_{3}$ when the wave packets 
$\psi_{1}(\mathbf{r}_{a})$ and 
$\psi_{2}(\mathbf{r}_{a})$ have again separated after passing through the region $I$, then 

3. The atoms arriving at D$_{1}$ (see figure 1) will have retained their energy and the
 cavity will be unexcited.
 
4. The atoms arriving at D$_{2}$ (see figure 1) will have lost their energy and the cavity 
will be excited.

These predictions are exactly the same as those obtained from standard quantum 
mechanics.  There are no observable differences between standard quantum mechanics and
the Bohm approach nor can there be simply because the Bohm approach
uses the same wave functions and the same formalism as is used in the usual approach
and therefore both approaches must end up with exactly the same probabilities. Because we are
using in addition the one-parameter solutions of the real part of the Schr\"{o}dinger equation given
in (19) we get a further insight into quantum phenomena.   

The Bohm way of looking at
the solutions avoids some of the difficulties presented by the standard approach.  It
has no ambiguity in the role played by the `quantum particle'.  It has no measurement
problem and definitely no schizophrenic cats.  It offers an explanation of the
behaviour of the probability currents discussed in section 3 and it also clearly shows
why the cavity cannot be used as a WW device.  It also draws attention to the
non-separability and non-locality aspects of systems described by entangled wave
functions, a feature that Bohr was content to describe as the 
inherent `wholeness of
the phenomenon'.

In spite of these positive aspects a word of caution should is
necessary.  There is no way to confirm that the quantum process is actually a particle
travelling along a path which is separate from the $R$ and $S$ fields.  These fields are
not classical fields that are to be included in a Hamiltonian and are not independent
of the particle itself.  They have a very different quality from the fields that arise in
classical physics and this suggests that there is a radically novel 
ontological process involved
that the Bohm approach captures in some coarse-grained approximation. 

Bohm and Hiley\cite{BH93}\cite{BH87}
 have suggested that these new fields can be understood in terms of the
new concept of active information, a concept that has been discussed elsewhere
(Maroney and Hiley \cite{MH}, Hiley \cite{BJH99} and Hiley and Pylkk\"{a}nen 
\cite{HP}).  In spite of these
limited successes, the nature of this deeper process is still very illusive and arises
essentially from the non-commutative structure of the quantum algebra.

In quantum
mechanics these non-commutative algebras have a profound implication for the nature of
spacetime, as is very clear from some of the attempts to understand non-commutative
geometry (Connes \cite{C}).  Some of these ideas have been applied to the EPR paradox
offering a possible understanding of non-locality through a deeper understanding of
the properties of spacetime (Grib and Zapatrin \cite{GZ}, Hiley 
\cite{BJH91}, Hiley and Fernandes \cite{HF}, Heller and Sasin \cite{HS}).  These papers move the discussion about the nature of
quantum non-separability and quantum non-locality on to a new level away from the
somewhat sterile discussions on the interpretation of the non-relativistic quantum
formalism.


\section{References}\label{sec.7}
\begin{thebibliography}{99}
\bibitem{Bohm52}
D. Bohm, 1952, {\em A Suggested Interpretation of the Quantum Theory in Terms of ÔHidden
 VariablesÕ}, Phys. Rev., {\bf 85}, 166-179;  {\bf 85}, 180-193.
\bibitem{BH87}
D. Bohm and B. J. Hiley, 1987, {\em An Ontological Basis for the Quantum Theory:
 I-Non-relativistic Particle Systems}, Phys. Reports {\bf 144}, 323-348. 
\bibitem{BH93}
D. Bohm and B. J. Hiley, 1993, {\em The Undivided Universe: an Ontological Interpretation 
of Quantum Theory}, Routledge, London.
\bibitem{PH}
P. Holland, 1993, {\em The Quantum Theory of Motion}, Cambridge University Press, Cambridge.
\bibitem{DGZ}
D. D\"{u}rr, S. Goldstein and N. Zanghi, 1992, {\em Quantum Equilibrium and the Origin of
 Absolute Uncertainty}, J. Stat. Phys., {\bf 67}, 843-907.
 \bibitem{PAU}
W. Pauli, 1953, {\em Remarques sur le probl\`{e}m des param\`{e}tres cachŽs dans la mŽchanique
 quantque et sur la thŽorie de lÕonde pilote, in Louis de Broglie, Physicien et
Penseur}, ed A. George, 3-42, \`{E}ditions Albin Michel, Paris.
\bibitem{Z}
H. D. Zeh, 1998, {\em Why Bohm's Quantum Theory?} quant-phy/9812059.
\bibitem{ESSW2}
B-G. Englert, M. O. Scully, G. Sussman and H. Walther, (ESSW2), 1992, 
{\em Surrealistic 
Bohm Trajectories},  Z. Naturforsch. {\bf 47a}, 1175-1186.
\bibitem{S}
M. O. Scully, 1998, {\em Do Bohm trajectories always provide a trustworthy physical
 picture of particle motion?} Phys. Scripta, {\bf T76}, 41-46.
 \bibitem{DHS}
C. Dewdney, L. Hardy and E. J. Squires, 1993, {\em How late measurements of quantum 
trajectories can fool a detector}, Phys. Lett. {\bf 184A}, 6-11.
 \bibitem{DF}
D. D\"{u}rr, W. Fusseder, S. Goldstein and N. Zanghi, 1993, {\em Comment on
``Surrealistic
 Bohm Trajectories"}, Z. Naturforsch. {\bf 48a}, 1261-1262.
\bibitem{ESSW3}
B-G. Englert, M. O. Scully, G. Sussman and H. Walther, (ESSW3), 1993, 
{\em Reply to Comment
 on ``Surrealistic Bohm Trajectories}, Z. Naturforsch. {\bf 48a}, 1263-1264.
\bibitem{NB61}
N. Bohr, 1961, {\em Atomic Physics and Human Knowledge},  Science Editions, New York.
\bibitem{AV}
Y. Aharonov and L. Vaidman, 1996, in {\em Bohmian Mechanics and Quantum Theory: an
Appraisal}, ed J. T. Cushing, A. Fine and S. Goldstein, Boston Studies in the
Philosophy of Science, {\bf 184}, 141-154, Kluwer, Dordrecht.
\bibitem{RG}
R. B. Griffiths, 1999, {\em Bohmian mechanics and consistent histories}, Phys. Lett., {\bf A261}, 227-234.
\bibitem{NB61B}
N. Bohr, 1961, {\em Atomic Physics and Human Knowledge}, p. 39, Science Editions, New York.
\bibitem{BJH99}
B. J. Hiley, 1999, {\em Active Information and Teleportation}. In { \em Epistemological
and Experimental Perspectives on Quantum Physics}, ed. D. Greenberger et al., 113-126,
Kluwer, Netherlands.
\bibitem{EPR}
A. Einstein, B. Podolsky, and N. Rosen, 1935, {\em Can Quantum-Mechanical Description of
 Physical Reality be Considered Complete?} Phys. Rev., {\bf 47}, 777-80.
 \bibitem{LL}
L. D. Landau and E. M. Lifshitz, 1977, {\em Quantum Mechanics}, p. 2 Pergamon Press, Oxford.
\bibitem{NB61A}
N. Bohr, 1961, {\em Atomic Physics and Human Knowledge}, p. 51, Science Editions, New York.
\bibitem{WH}
W. Heisenberg, 1927, {\em \"{U}ber den anschaulichen Inhalt der quantumtheoretischen Kinematik
und Mechanik},  Zeit. fŸr Phys. {\bf 43}, 172-98.  English translation in J. A. Wheeler and
W. Zurek, 1981, {\em Quantum Theory and Measurement}, p. 63-5.
\bibitem{NB51}
N. Bohr 1951, in {\em Albert Einstein: Philosopher-Physicist} ed, P. A. Shilpp, p. 222, 
Tudor, New York.
\bibitem{WH58}
W. Heisenberg, 1958, {\em Physics and Philosophy}, Harper-Row, New York,
Or {\em The representations of the nature of contemporary physics}, Daedalus, 
{\bf 87}, 95-108.
\bibitem{SEW}
M. O. Scully, B.-G. Englert and H. Walther,1991, {\it Quantum Optical Tests of
Complementarity}, Nature, {\bf 351},111-116.
\bibitem{NB64}
N. Bohr, 1961, {\em Atomic Physics and Human Knowledge}, p. 64, Science Editions, New York.
\bibitem{JW}
J. A. Wheeler, 1996, {\em At Home in the Universe}, p. 344, AIP, New York.
\bibitem{NBPR}
N. Bohr, 1935,{\em Can quantum-mechanical description of physical reality be considered complete?}, Phys.
Rev., {\bf 48}, 696-702.
\bibitem{JB64}
 J. S. Bell, 1964, {\em On the Einstein-Podolsky-Rosen paradox}, Physics, 
{\bf 1}, 195-200.
\bibitem{JB66}
J. S. Bell, 1966, {\em On the problem of hidden variables in quantum theory}, Rev. Mod. Phys.
 {\bf 38}, 447-452.
 \bibitem{DPH}
C. Dewdney, C. Philippidis and B. J. Hiley, 1979, {\em Quantum Interference and the 
Quantum Potential},  Il Nuovo Cimento {\bf 52B}, 15-28.
 \bibitem{BH75}
D. Bohm and B. J. Hiley, 1975,  {\em On the Intuitive Understanding of Nonlocality as 
Implied by Qauntum Theory},   Found. Phys.,  {\bf 5},  93-109.
\bibitem{MH}
O. Maroney and B. J. Hiley, 1999, {\em Quantum State Teleportation understood through the
 Bohm Interpretation},  Found. Phys. {\bf 29}, 1403-15.
 \bibitem{BHK}
D. Bohm, B. J. Hiley and P.N. Kaloyerou, 1987, {\em An Ontological Basis for the Quantum Theory:
II -A Causal Interpretation of Quantum Fields}, Phys. Reports {\bf 144}, 349-375. 
\bibitem{K}
P. N. Kaloyerou, 1994, {\em The Causal Interpretation of the Electromagnetic Field}, 
Phys. Rep. {\bf 244}, 287-385.
\bibitem{LD} M. M. Lam and C. Dewdney, {\it The Bohm Approach to Cavity Quantum Scalar
Field Dynamics, I and II}, Found. Phys. {\bf 24}, 3-60, (1994).
 \bibitem{BH93A}
D. Bohm and B. J. Hiley, 1993a, {\em The Undivided Universe: an Ontological Interpretation
 of Quantum Theory}, p. 29, Routledge, London.
 \bibitem{BH81}
D. Bohm and B. J. Hiley, 1981, {\em Non-locality in Quantum Theory understood in Terms of 
Einstein's Non-linear Field Approach}, Found. Phys. {\bf 11}, 529-546.
 \bibitem{HP}
B. J. Hiley and P. Pylkk\"{a}nen, 1997, {\em Active Information and Cognitive Science-A Reply 
to Kiesepp\"{a}}, in Brain, Mind and Physics, ed. P. Pylkk\"{a}nen, P. 
Pylkk\"{o}, and A.
HautamŠki, 64-85, IOS Press, Amsterdam.
\bibitem{C}
A. Connes, 1994, {\em Noncommutative Geometry}, Academic Press, San Diego.
 \bibitem{GZ}
A. A. Grib and R. R Zapatrin, 1992, {\em Topology Lattice as Quantum Logic}, Int. J. Theo.
 Phys. {\bf 31}, 1093-1101.
\bibitem{BJH91}
B. J. Hiley,  1991,  {\em Vacuum or Holomovement}, in {\em The Philosophy of Vacuum},
ed., S.
 Saunders and H. R. Brown, pp. 217-249,  Clarendon Press, Oxford.
 \bibitem{HF}
B. J. Hiley and M. Fernandes, 1997, {\em Process and Time}, in {\em Time, Temporality
and Now},
 ed., H. Atmanspacher and E. Ruhnau, 365-383,  Springer-Verlag.
 \bibitem{HS}
M. Heller and W. Sasin, 1998, {\em Einstein-Podolsky-Rosen Experiment from Non-commutative 
Quantum Gravity}, Particles, Fields and Gravitation, p. 234, AIP.
\end{thebibliography}
\end{document}